\documentclass[twocolumn, english,showpacs,preprintnumbers,amsmath,amssymb]{revtex4}
\usepackage{amsmath}
\usepackage{amssymb}
\usepackage{graphicx}
\usepackage{babel}
\usepackage{times}

\allowdisplaybreaks[3]

\graphicspath{{Figures/}}

\begin{document}

\title{Quantum spin liquid with a Majorana Fermi surface on the three-dimensional hyperoctagon lattice}

\author{M. Hermanns}

\author{S. Trebst}

\affiliation{Institute for Theoretical Physics, University of Cologne, 50937 Cologne, Germany}

\date{\today}

\begin{abstract}
Motivated by the recent synthesis of $\beta$-Li$_2$IrO$_3$ --  a spin-orbit entangled $j=1/2$ Mott insulator with a three-dimensional lattice structure of the Ir$^{4+}$ ions -- we consider generalizations of the Kitaev model  believed to capture some of the microscopic interactions between the Iridium moments on various trivalent lattice structures in three spatial dimensions. 
Of particular interest is the so-called hyperoctagon lattice -- the premedial lattice of the hyperkagome lattice, for which the ground state is a gapless quantum spin liquid where the gapless Majorana modes form an extended two-dimensional Majorana Fermi surface. 
We demonstrate that this Majorana Fermi surface is inherently protected by lattice symmetries and discuss possible instabilities. 
We thus provide the first example of an analytically tractable microscopic model of interacting SU(2) spin-1/2 degrees of freedom in three spatial dimensions that harbors a spin liquid with a two-dimensional spinon Fermi surface. 
\vspace{5mm}
\end{abstract}


\maketitle


\section{Introduction}
\label{Introduction}

Frustrated quantum magnets can exhibit highly unconventional ground states, in which local moments are highly correlated but 
nevertheless evade a conventional ordering transition and remain strongly fluctuating down to zero temperature. These unusual
states are commonly referred to as quantum spin liquids \cite{SpinLiquids} -- despite their rather diverse physical properties ranging from gapped states with an emergent topological order to gapless states with an emergent spinon Fermi surface.
A common motif in the search for quantum spin liquids has been to look for quantum antiferromagnets on geometrically frustrated lattices, 
i.e. lattices where the elementary building blocks prohibit the formation of a conventional N\'eel state. Paradigmatic examples of 
geometric frustration include lattices formed by corner-sharing tetraedra such as the pyrochlore lattice, or by corner-sharing triangles such as the kagome lattice in two spatial dimensions and the hyperkagome lattice in three spatial dimensions.
An alternative route to induce frustration in a quantum magnet is to look for systems, in which competing interactions cannot be 
simultaneously satisfied. Archetypal examples of such exchange frustration are given by the quantum compass models \cite{CompassModels},
in which the easy-axis of an anisotropic spin exchange strongly depends on the spatial orientation of the exchange path -- a scenario which can 
prohibit even a ferromagnet on a bipartite lattice from undergoing a finite-temperature ordering transition. The best known example
in this class of compass models is the Kitaev model \cite{KitaevModel} on the honeycomb lattice, in which the easy-axis of an Ising-like 
spin exchange points along the $x$, $y$, and $z$ directions for the three different bond types of the hexagonal lattice, which is captured by
the Hamiltonian 
\begin{equation}
  H_{\rm Kitaev} = \sum_{\gamma \rm-links} J_\gamma \sigma_i^\gamma \sigma_j^\gamma \,,
  \label{eq:Kitaev}
\end{equation}
where SU(2) spins $\sigma$ on sites $i$ and $j$ are connected via a bond in the $\gamma = x,y,z$ direction.
The Kitaev model is quintessential in that it harbors three different types of quantum spin liquids -- a gapped, $\mathbb{Z}_2$ topological spin liquid if one of the three exchange couplings is significantly larger than the couplings associated with the two other bond directions (i.e. $J_z > 2 J_{x,y}$), and a gapless spin liquid in the vicinity of equal-strength exchange couplings ($J_x \approx J_y \approx J_z$). If applying an external magnetic field along the $111$-direction, the latter can be gapped out into a topological spin liquid with non-Abelian vortex excitations. 
The Kitaev model not only stands out for the unusual richness of its ground states, but the fact that it is one of the very few examples of an interacting spin model that can be rigorously solved.
It should, however, be pointed out that the Kitaev model has not only attracted the imagination of phenomenologically inclined theorists, but has also stirred some excitement in the materials oriented community after it has been pointed out that the significantly enhanced spin-orbit coupling in 5d transition metal oxides and in particular certain Iridates can give rise to unconventional Mott insulators where the local moment is a spin-orbit entangled $j=1/2$ moment
\cite{Sr2IrO4a,Sr2IrO4b}. 
The orbital contribution to these moments results in a highly anisotropic, spatially oriented exchange \cite{Jackeli09}, 
which can in fact mimic those of the Kitaev model \eqref{eq:Kitaev}.
In terms of actual materials the layered Iridates Na$_2$IrO$_3$ and Li$_2$IrO$_3$ have attracted much recent interest and are intensely discussed 
\cite{Chaloupka10,Reuther11,Jiang11,Na2IrO3Thermodynamics,Na2IrO3Neutrons,Chaloupka13}
as possible candidate materials realizing the two-dimensional honeycomb Kitaev model.

In this manuscript, we turn to generalizations of the Kitaev model on three-dimensional lattices -- a move that is prompted by the recent synthesis of $\beta$-Li$_2$IrO$_3$ \cite{hyperTakagi}, which forms a truly three-dimensional lattice structure of the Ir$^{4+}$ ions. This structure, which has quickly been dubbed hyperhoneycomb lattice \cite{hyperTakagi}, keeps the trivalent vertex structure of the hexagonal lattice and thereby the essential feature allowing for an analytical solution of the Kitaev model. In fact, the Kitaev model on the hyperhoneycomb lattice had been identified and studied before by Mandal and Surendran \cite{Mandal09} who reported the occurrence of a gapless spin liquid with an emergent spinon Fermi surface on a line in momentum space for approximately equal-strength interactions ($J_x \approx J_y \approx J_z$) as well as the occurrence of a gapped topological spin liquid for anisotropic exchange strength \cite{Mandal11}. More recently, extensions to a Heisenberg-Kitaev model \cite{Chaloupka10} have  established the stability of
this gapless phase in the presence of weak isotropic spin exchange \cite{hyperKim,hyperKimchi,hyperKim2}.

\begin{figure}[t]
\includegraphics[width=\columnwidth]{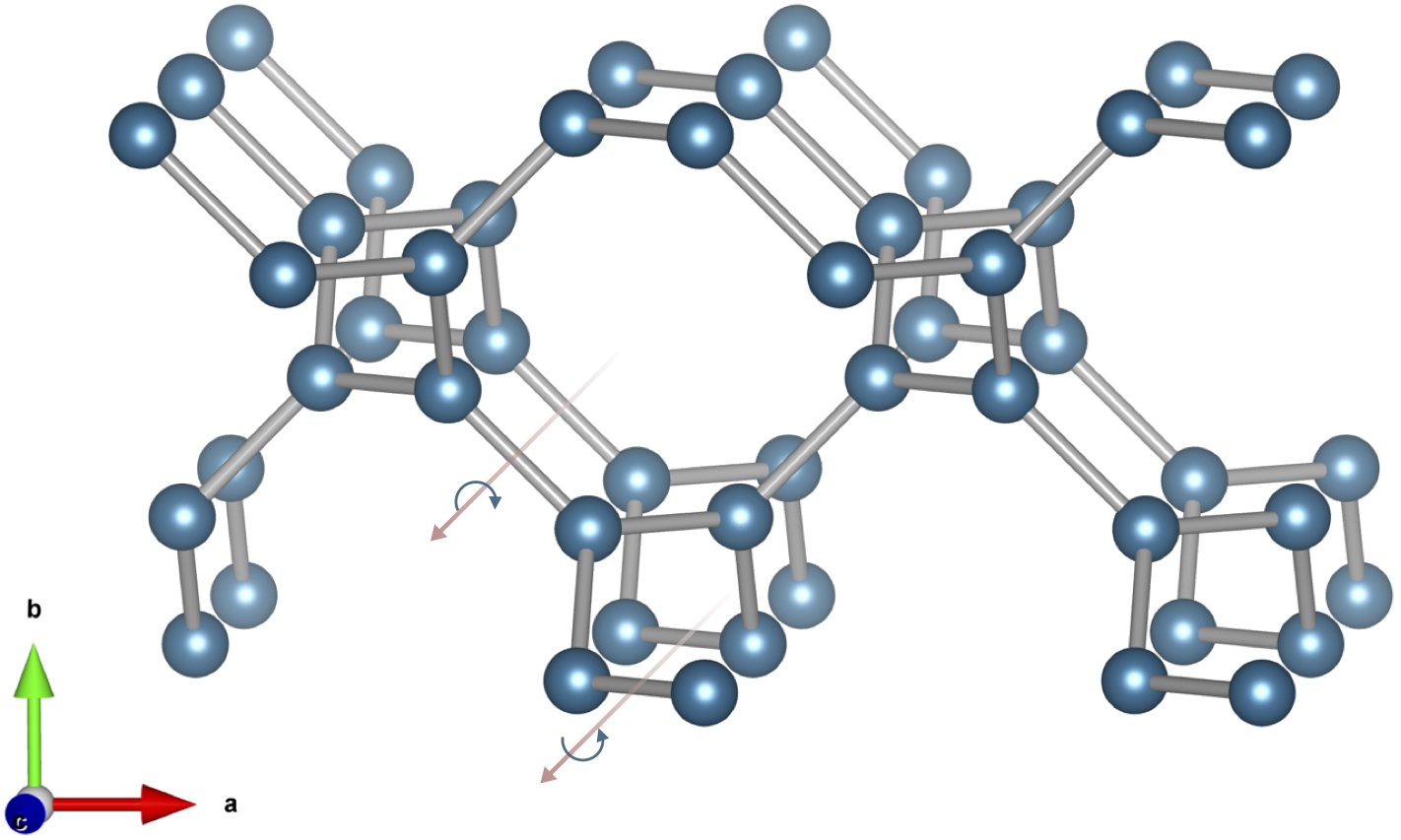}
\caption{(color online) Illustration of the hyperoctagon lattice, a trivalent structure which contains two elementary motifs 
  	      -- a spiraling octagonal helix and a counter-spiraling square helix. }
\label{fig:HyperoctagonLattice}
\end{figure}

\begin{figure}[t]
\includegraphics[width=\columnwidth]{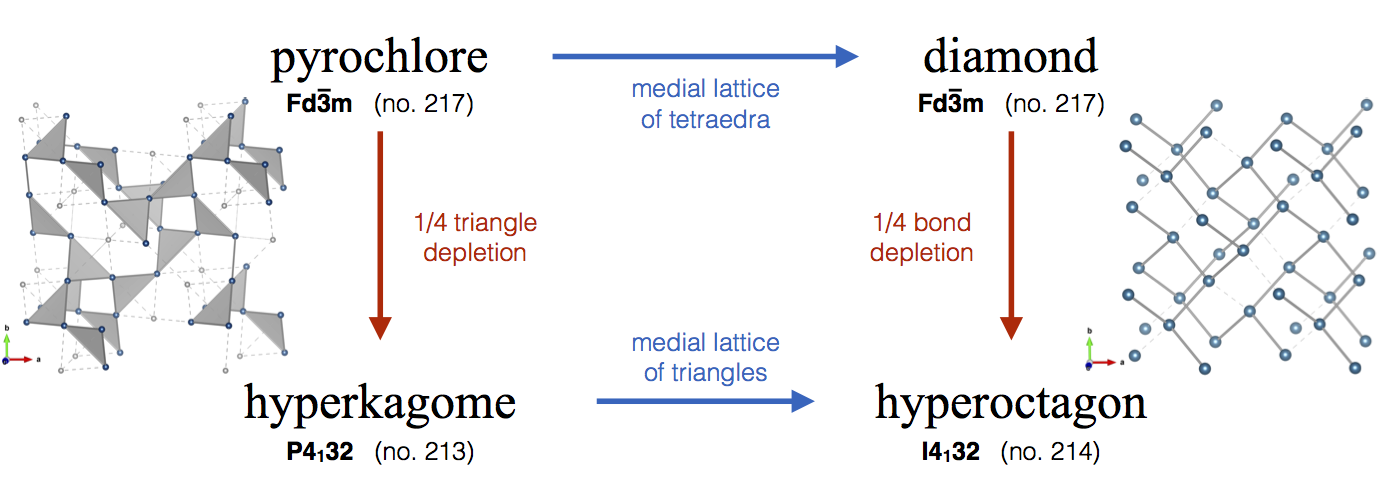}
\caption{(color online) Relation between various three-dimensional lattices.
 	      The hyperoctagon lattice is the premedial lattice of the hyper\-kagome lattice, which can be 
	      obtained  from the pyrochlore lattice via depletion of 1/4 of the triangles. 
	      The premedial lattice of the pyrochlore is the diamond lattice, which can be depleted by 1/4
	      of its bonds to obtain the hyperoctagon lattice.}
\label{fig:LatticeSummary}
\end{figure}

This motivated us to ponder alternative three-dimensional lattices that keep the trivalent vertex structure and led us to consider what we call the hyperoctagon lattice
\cite{HyperoctagonFiles}
illustrated in Fig.~\ref{fig:HyperoctagonLattice}. The hyperoctagon lattice is closely related to the hyperkagome lattice -- the hyperoctagon lattice is the premedial lattice of the hyperkagome lattice obtained by shrinking each triangle of the 
hyperkagome lattice to a single vertex and the new bonds indicating the original connectivity of the triangles,
schematically summarized in Fig.~\ref{fig:LatticeSummary}. 
The hyperoctagon lattice is a chiral lattice, which contains two elementary motifs -- a spiraling octagonal helix and a counter-spiraling 
square helix as illustrated in Fig.~\ref{fig:HyperoctagonLattice}.
Its space group I4$_1$32 (no. 214) indicates the presence of 4-, 3- and 2-fold (skew) symmetries (the details of which we will provide 
below) that will turn out to play an essential role in stabilizing the gapless modes of the quantum spin liquid emerging for the Kitaev
model on this lattice. The presence of these symmetries is also the key distinction to the hyperhoneycomb lattice, another somewhat 
less symmetric three-dimensional trivalent lattice structure which has been revealed in the recent synthesis of $\beta$-Li$_2$IrO$_3$.

Our main result is the observation of a gapless quantum spin liquid with an extended two-dimensional Majorana Fermi surface around the point of isotropic 
couplings for the Kitaev model on the hyperoctagon lattice. This result is rigorously established by an exact analytical solution of the spin model, 
which can be cast into a free fermion system by Majorana fermionization, thus employing the same powerful techniques that have 
already allowed the solution of the Kitaev model on other trivalent lattices \cite{KitaevModel,Mandal09,ChiralSpinLiquid}.

Our discussion in the remainder of the paper is structured as follows. 
In Section \ref{sec:lattice} we will discuss trivalent lattice structures in two and three spatial dimensions and in particular
provide a detailed introduction of the hyperoctagon lattice.
The Kitaev model on the hyperoctagon lattice is subsequently introduced and exactly solved in Section \ref{sec:kitaev_model}
where we also provide a detailed discussion of its ground state phase diagram, in particular the gapless spin liquid 
with a Majorana Fermi surface emerging for a broad range of parameters. 
Possible instabilities of the Majorana Fermi surface are discussed in Section \ref{sec:FermiSurface}.
We conclude with an outlook in Section \ref{sec:outlook}.

\section{The hyperoctagon lattice}
\label{sec:lattice}

\begin{figure}[b]
\includegraphics[width=\columnwidth]{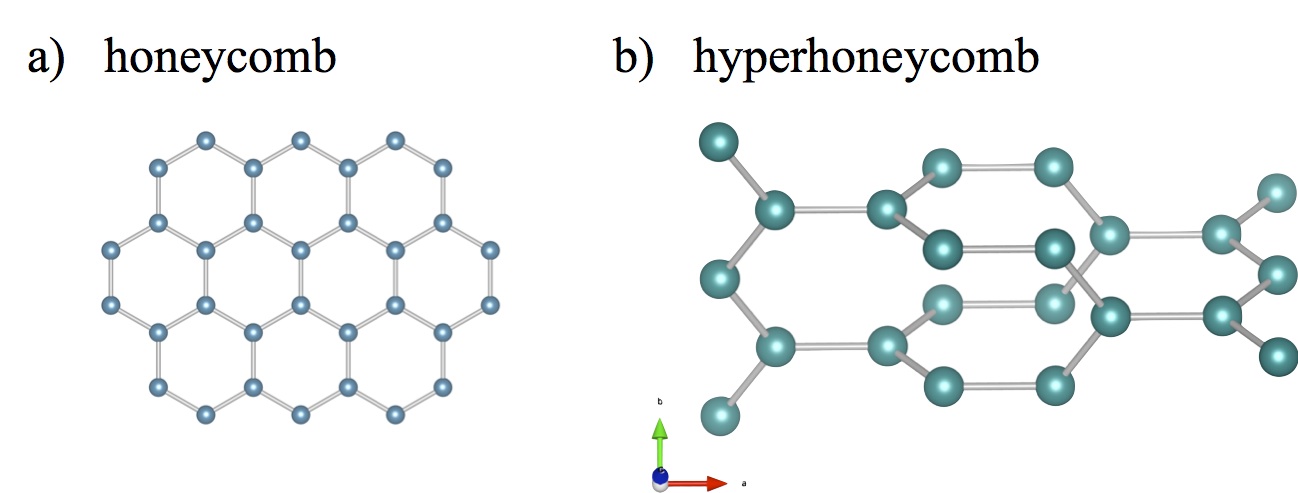}
\caption{(color online) Illustration of a) the honeycomb and b) the hyperhoneycomb lattice. 
}
\label{fig:honeycombs}
\end{figure}

Before we dive into the physics of the Kitaev model we start our discussion with a short review of the underlying lattice structure.
In its original form the Kitaev model has been discussed for the honeycomb lattice, a two-dimensional lattice with a trivalent 
coordination of all vertices as depicted in Fig.~\ref{fig:honeycombs} a). 
Keeping this motif of a trivalent lattice structure the model can readily be associated with a broader class of lattices -- a move that
not only allows a $1:1$ assignment of the three different exchange types to the bonds around the vertices, but is also key to keep
the analytical tractability of the model, which we will review in the following section
\cite{FootnoteOtherLattices}.
In two spatial dimensions one such generalization is the square-octagon lattice
of Ref.~\onlinecite{SquareOctagonModel}. 
In three spatial dimensions such trivalent lattices are considerably less common. One example is the so-called hyperhoneycomb lattice,
which is depicted in Fig.~\ref{fig:honeycombs} b). The elementary building blocks of the hyperhoneycomb lattice are zig-zag chains 
running along the crystallographic $b$ and $c$ axis, respectively, as depicted in Fig.~\ref{fig:honeycombs} b). 
These zig-zag chains are coupled by bonds along the $a$ axis, readily implying that there is no general symmetry possibly 
interchanging the three crystallographic axises.
Another example of a trivalent lattice in three dimensions is the so-called hyperoctagon lattice, which we describe in detail in the following.

\subsection{Lattice symmetries}
\label{sec:lattice_symmetries}
\begin{figure}[t]
\includegraphics[width=\columnwidth]{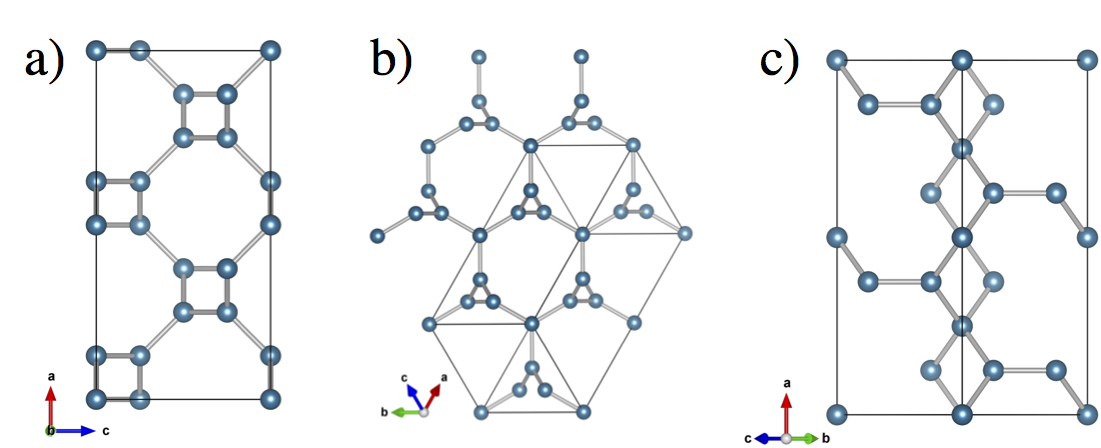}
\caption{(color online) Visualization of the (screw-) rotation symmetries of the hyperoctagon lattice,
	      which belongs to the cubic space group I$4_132$ (no. $214$).  
a) View along (0,1,0) as an example of  a 90 degree screw-rotation symmetry. 
b) View along (1,1,-1) as an example of a 120 degree rotation symmetry. 
c) View along (0,1,1) as an example of a 180 degree rotation symmetry. 
}
\label{fig:rotations}
\end{figure}

The hyperoctagon lattice is a body-centered cubic lattice without inversion symmetry. Its symmetries correspond to space group I4$_1$32 (no. 214). In cartesian coordinates, the atomic positions can be constructed starting from the point $\frac 1 8  (1,1,1)$ in the unit cell and applying all symmetry transformations of the space group on it. In particular, the symmetries of space group I4$_1$32 include the following: 
i) a four-fold symmetry which is obtained by 90 degree \emph{screw}-rotations around the (1,0,0), (0,1,0), or (0,0,1) directions. 
ii) a three-fold symmetry which leave the lattice invariant under 120 degree rotations around the (1,1,1), (-1,1,1), (1,-1,1), or (1,1,-1) directions,
and 
iii) a two-fold symmetry corresponding to 180 degree rotations around the directions $(\pm 1,  1,0)$ , $(\pm 1, 0, 1)$, and $(0, \pm  1, 1)$. 
As a guide to the eye, Fig.~\ref{fig:rotations} shows the projection of the lattice onto the  planes normal to the rotation axis for three examples of the above (screw-)rotations. Note that the (projected) square-octagon structure in Fig.~\ref{fig:rotations} a) is not planar, which is why the additional translation is needed.  

\subsection{Possible materials}

While the hyperoctagon lattice arises quite naturally as the premedial lattice of the hyperkagome lattice, which is realized for instance  in the spin liquid material Na$_4$Ir$_3$O$_8$ \cite{Hyperkagome}, there are so far no known realizations of the hyperoctagon lattice in the diverse family of recently synthesized Iridates. To provide some abstract guidance as to which chemical compositions might possibly realize magnetic  hyperoctagon systems, we have made an attempt at designing possible materials candidates.

\begin{figure}[t]
\includegraphics[width=\columnwidth]{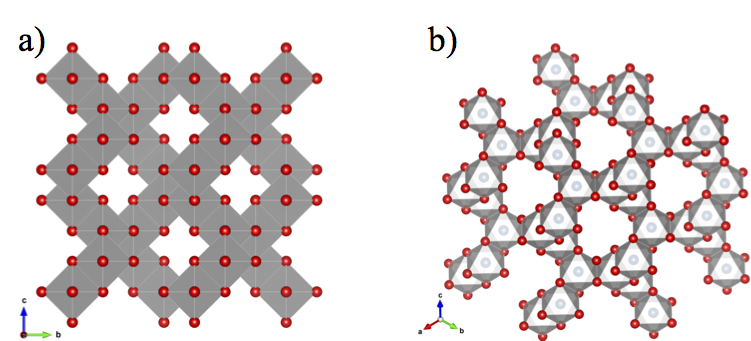}
\caption{(color online)  Structure of the edge sharing O$_6$ octahedra around the central Ir atoms. a) View along the (1,0,0) direction. b) View along the (1,1,1) direction.}
\label{fig:O6cages}
\end{figure}

With the Iridium atoms assumed to occupy the sites of the hyperoctagon lattice, our further thinking is guided by the microscopic prerequisites that allow a dominant anisotropic Kitaev-like interaction to emerge -- the occurrence of double Ir-O-Ir exchange paths that suppress the isotropic spin exchange \cite{Jackeli09,Chaloupka10}. The latter can be achieved by placing the Iridium atoms in bond-sharing IrO$_6$ cages.
In fact, the symmetries of the hyperoctagon lattice allow to embed each Iridium atom into a perfectly undistorted IrO$_6$ octahedron, when 
placing the oxygen atoms at position $1/8 (1,-1,1)$ in the unit cell. The resulting IrO$_3$ structure is illustrated in Fig.~\ref{fig:O6cages}.
Such a sparse octahedron structure has indeed been observed for the subhalides, e.g. La$_3$Br$_3$Si \cite{La3X3Z}, where the Silicium atoms form the hyperoctagon lattice and the Lanthanum atoms form octahedra around them.

Finally, one might want to attempt to fill the remaining interstitial sites of the octahedron structure. Taking into account the space group
symmetries there are several distinct ways of doing so, as described in some detail in Appendix \ref{sec:AppSpaceGroupMaterials}.
In particular, one might start to add atoms to a single interstitial site (and its space group related siblings) as illustrated in Fig.~\ref{fig:space_group}~a) of the Appendix. This would result in the chemical composition of the alkaline-earth iridates AIrO$_3$ where A is one of the alkaline-earth elements Ca, Sr, or Ba. The alkaline-earth iridates are known to exhibit quite distinct electronic properties for the different A-site materials, including an $S=1/2$ antiferromagnetic Mott insulator for CaIrO$_3$ \cite{CaIrO3}, a weak ferromagnetic semiconductor for BaIrO$_3$ \cite{BaIrO3} and a non-Fermi liquid metal in SrIrO$_3$ \cite{SrIrO3}. While  various crystal structures have been reported for the different AIrO$_3$ compounds, no crystals in space group I$4_132$  have so far been synthesized for any of the alkaline-earth iridates.
An alternative possibility to fill the interstitial sites is to add two additional atoms resulting in a chemical composition of the form A$_2$IrO$_3$ with A being one of the alkali metals Na or Li. The resulting crystal structure is illustrated in Fig.~\ref{fig:space_group}~b) of the Appendix. This is a particularly interesting idea to entertain as it would point to a possible existence of a third crystallization pattern for A$_2$IrO$_3$ beyond the already known examples of  quasi two-dimensional honeycomb layers and the recently synthesized three-dimensional hyperhoneycomb structure. 

\section{Kitaev model}
\label{sec:kitaev_model}

Not only motivated by a possible relevance to future materials, but also driven by a curiosity to explore unusual spin liquid states 
we now turn to a three-dimensional variant of the  Kitaev model on the hyperoctagon lattice.
We proceed with an introduction and precise definition of the model and discuss some of its general properties before presenting an
analytical solution of the model in terms of an exact Majorana fermionization of the spin degrees of freedom. Finally, we present our 
main result of identifying a gapless spin liquid ground state with a Majorana Fermi surface.


\subsection{The model}
\label{sec:the_model}
Kitaev originally introduced his elementary spin model as a system of SU(2) spin-1/2 degrees of freedom interacting on the  two-dimensional honeycomb lattice.  
Its fundamental beauty not only arises from its exact analytical solution, but the fact that the spin model harbors a number of paradigmatic ground states -- besides an Abelian topological phase, it exhibits an extended gapless spin liquid ground state, which can be gapped out into a non-Abelian topological phase by an external magnetic field.
This variety of different ground states arises from highly frustrated spin interactions which favor the alignment (or anti-alignment) of different spin components along the three principle directions of the honeycomb lattice.

\begin{figure}[t]
\includegraphics[width=\columnwidth]{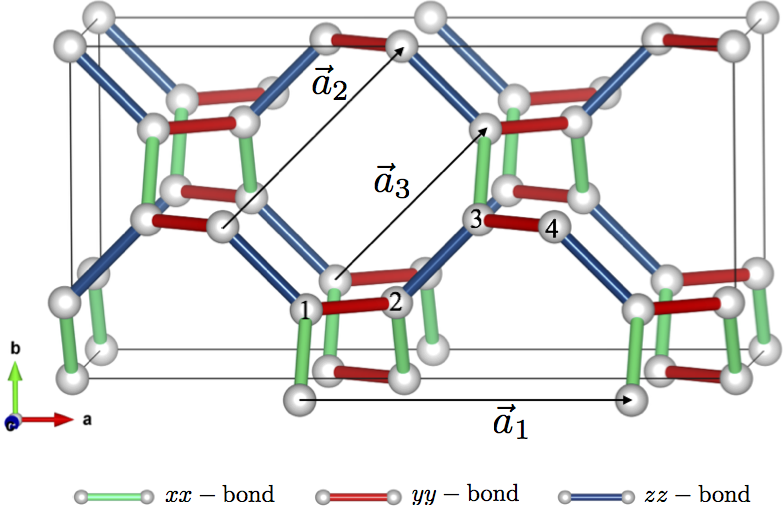}
\caption{(color online) Illustration of the different couplings of the Kitaev model on the hyperoctagon lattice. 
              Green bonds correspond to $xx$-couplings, red bonds to $yy$-couplings, and blue bonds to $zz$-couplings, respectively.}
\label{fig:KitaevModel}
\end{figure}

Here we generalize this idea to the three-dimensional hyper\-octagon lattice, which because of its trivalent vertices allows the definition of an analogous spin model. To this end, we cover the lattice with bonds that favor spin alignments along the $x$, $y$, and $z$ directions and which we call $xx$, $yy$, and $zz$-bonds, respectively. While there are many different ways to realize such coverings on a given lattice, 
there is only a single covering that is compatible with all the lattice translation symmetries of the hyperoctagon lattice. This unique covering is illustrated in Fig.~\ref{fig:KitaevModel} and will serve as our primary definition of the Kitaev model on this lattice. We will briefly discuss alternative models based on other coverings in Section \ref{sec:FermiSurface}.

In order to provide a self-contained description of the model, we start by introducing the four-site unit cell compatible with the unique
covering of exchange bonds such that all lattice translation symmetries are kept. The atomic positions in this unit cell are given by
\begin{align}
\label{eq:unit_cell}
\mathbf r_1&=\mathbf R+\frac 1 8 (-3,-1,1)\nonumber\\
\mathbf r_2&=\mathbf R+\frac 1 8 (-1,-1,-1)\nonumber\\
\mathbf r_3&=\mathbf R+\frac 1 8 (1,1,-1)\nonumber\\
\mathbf r_4&=\mathbf R+\frac 1 8 (3,1,1), 
\end{align} 
where $\mathbf R=(\frac 1 2, \frac 1 4, 0)+\sum_{j=1}^3 n_j \mathbf a_j$ is the unit cell position.
The offset in $\mathbf R$ is chosen to be consistent with the conventions in Section \ref{sec:lattice_symmetries} and can be mostly ignored in the following discussion. 
The corresponding lattice translation vectors are then given as 
\begin{align}
\label{eq:TranslVectors}
\mathbf a_1 &=(1,0,0)\nonumber\\
\mathbf a_2 &=\frac 1 2 (1,1,-1)\nonumber\\
\mathbf a_3 &=\frac 1 2 (1,1,1) \,,
\end{align}
which are also illustrated in Fig.~\ref{fig:KitaevModel}.

With these definitions in place we can now define the Kitaev Hamiltonian on the hyperoctagon lattice as 
\begin{align}
H&=-\sum_{\mathbf{R}} 
J_{x}\left(\sigma_{1}^{x}(\mathbf{R})\sigma_{3}^{x}(\mathbf{R}-\mathbf{a}_{2}) +\sigma_{2}^{x}(\mathbf{R})\sigma_{4}^{x}(\mathbf{R}-\mathbf{a}_{3})\right) \nonumber\\
&+J_{y}\left(\sigma_{1}^{y}(\mathbf{R})\sigma_{2}^{y}(\mathbf{R})+\sigma_{3}^{y}(\mathbf{R})\sigma_{4}^{y}(\mathbf{R}) \right)\nonumber \\
&+J_{z}\left(\sigma_{2}^{z}(\mathbf{R})\sigma_{3}^{z}(\mathbf{R})+ \sigma_{1}^{z}(\mathbf{R})\sigma_{4}^{z}(\mathbf{R}-\mathbf{a}_{1})\right)\,.
\label{eq:SpinHamRealSpace}
\end{align}
For the following discussion, 
it is beneficial to introduce a `bond operator' $K_{i,j}$ for a bond $\langle i,j\rangle$
\begin{align}
\label{eq:bond_operator}
K_{i,j}&= \left\{\begin{array}{cc} \sigma_i^x \sigma_j^x & \mbox{if $\langle i,j\rangle $ is  of $xx$-type} \\
 \sigma_i^y \sigma_j^y & \mbox{if $\langle i,j\rangle $ is of $yy$-type} \\
  \sigma_i^z \sigma_j^z & \mbox{if $\langle i,j\rangle $ is of $zz$-type} 
  \end{array}\right.\, .
\end{align}
In terms of these bond operators, the Hamiltonian then reduces to the compact form introduced earlier
\begin{align}
 H &=\sum_{\gamma-{\rm links}} J_{\gamma} K_{i,j} .
\end{align}


\subsection{Loops and conserved quantities}
\label{ssec:loops}

\begin{figure}[t]
\includegraphics[width=\columnwidth]{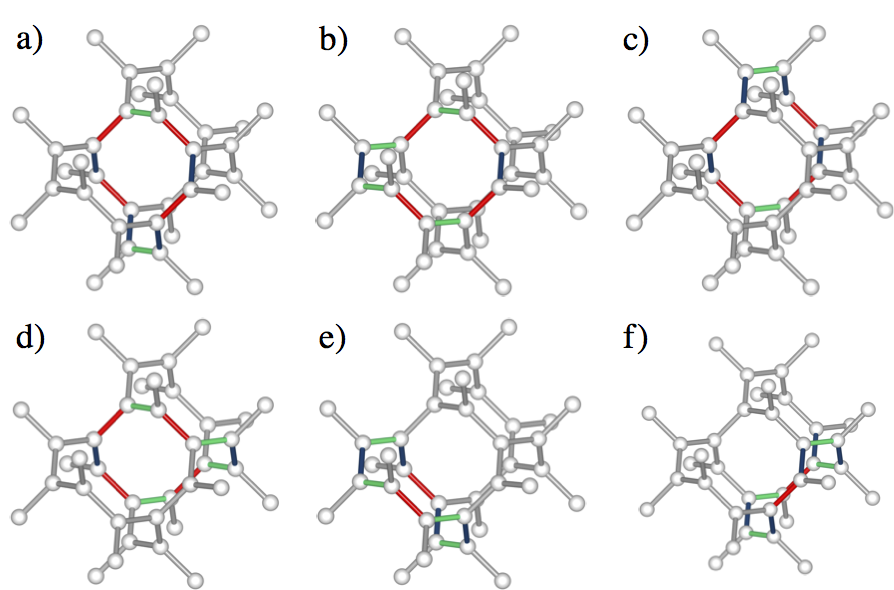}
\caption{(color online) 
	     The six distinct loops of the hyperoctagon lattice. 
	     Each loop contains ten bonds with two coupling types contributing four bonds and one coupling type contributing only two bonds.
	     Note that the six loops realize all possible combinations reflecting the symmetries of the hyperoctagon lattice. 
	     }
\label{fig:Loops_hyperoctagon}
\end{figure}

Our first step in analyzing Hamiltonian \eqref{eq:SpinHamRealSpace} is to identify conserved quantities, which we will find to be intimately connected to closed paths (or loops) on the lattice.
The elementary loops of the hyperoctagon lattice have length ten. For each unit cell there are six distinct such loops, which are visualized in  Fig.~\ref{fig:Loops_hyperoctagon}. All other elementary loops can be obtained by lattice translations. 
For each loop $l$ we can define a corresponding loop operator $W_l$ , which measures the `magnetic flux' through the plaquette that is enclosed by $l$. We can define the loop operator by the product of  bond operators of all the bonds contained in the loop
\begin{align}
\label{eq:loopOperator}
W_l &=\prod_{\langle i,j\rangle} K_{i,j} \,. 
\end{align}
Because of the even length of the loops these loop operators square to the identity, thus they have eigenvalues $\pm 1$. 
It can further be verified that the loop operators commute with the Hamiltonian \eqref{eq:SpinHamRealSpace} as well as with each other. 
Each loop operator thus defines an `integral of motion' and a corresponding conserved quantity -- the extensive number of which greatly simplifies the problem. For one, we can divide the Hilbert space into distinct sectors that are each labeled by the eigenvalues of all the loop operators $W_l$ and restrict the Hamiltonian to a particular sector. 

\begin{figure}[t]
\includegraphics[width=\columnwidth]{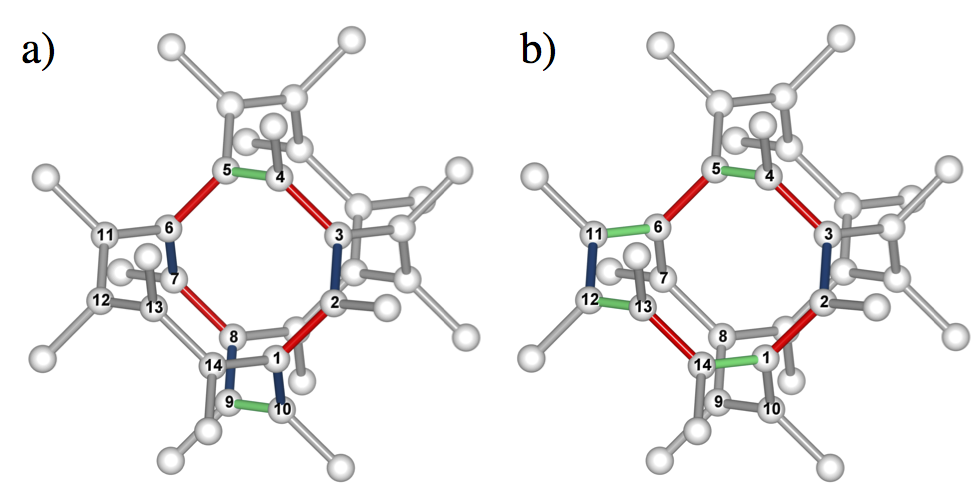}
\caption{(color online) 
	     Two examples of elementary loops. 
	     The sites are numbered  to facilitate the discussion in the main text. 
	     }
\label{fig:loops_numbered}
\end{figure}

Before proceeding we note that an alternative definition of the loop operators can be formulated as a product over all sites contained in the loop:
\begin{align}
\label{eq:loopOperator2}
\tilde{W}_l &=\prod_{i \in l} \sigma_i^{\gamma_i}\, ,
\end{align}
where $\gamma_i$ corresponds to the spin component at each vertex, which is not included along the loop.
For instance, the loop operator of the loop  in Fig.~\ref{fig:loops_numbered} a)  is given by
\begin{align}
\label{eq:loopa)}
\tilde{W}_{l_a}&=\sigma_1^x \sigma_2^x\sigma_3^x\sigma_4^z\sigma_5^z\sigma_6^x\sigma_7^x\sigma_8^x\sigma_9^y\sigma_{10}^y\, .
\end{align}
In both the honeycomb and hyperhoneycomb model, the two definitions of the loop operator, $W_l$ in \eqref{eq:loopOperator} and $\tilde W_l$ in  \eqref{eq:loopOperator2}, are identical. 
However, in our particular case there is a relative minus sign between the two, i.e. $ W_l  = -\tilde W_l$, which leaves some freedom in how to define the magnetic flux. 
In the following, we define  magnetic flux by the eigenvalue of \eqref{eq:loopOperator}: 
if $W_l$ has eigenvalue $-1$ we say that there is a magnetic flux (vortex)  penetrating the plaquette enclosed by $l$, while an eigenvalue $+1$ corresponds to no flux \cite{FootnoteLoopOperator}. 


\subsubsection*{Minimal volumes and flux sectors}
\label{sec:flux_sectors}
A further important difference to the two-dimensional case is that the loop operators are not all linearly independent. As an example, we  consider the two loops depicted in  Fig.~\ref{fig:loops_numbered}. The loop operator for the loop in Fig.~\ref{fig:loops_numbered}~a) is given in Eq. \eqref{eq:loopa)}, while the one for the loop in Fig.~\ref{fig:loops_numbered}~b) is given by
\begin{align}
W_{l_b}&=-\sigma_1^z \sigma_2^x\sigma_3^x\sigma_4^z\sigma_5^z\sigma_6^z\sigma_{11}^y\sigma_{12}^y\sigma_{13}^z\sigma_{14}^z \, . 
\end{align}
Note that we can define a third loop of length ten by combining the bonds that are contained in $W_{l_a}$ or $W_{l_b}$, but not in both of them (see the loop illustrated in panel e) of Fig.~\ref{fig:Loops_hyperoctagon})
\begin{align}
W_{l_c}&=-\sigma_6^y \sigma_7^x \sigma_8^x\sigma_9^y \sigma_{10}^y\sigma_1^y\sigma_{14}^z\sigma_{13}^z\sigma_{12}^y\sigma_{11}^y\, . 
\end{align}
The product of the three loops is the identity operator
\begin{align}
W_{l_a}W_{l_b}W_{l_c}&=(-1)^3\sigma_6^x \sigma_6^z \sigma_6^y \sigma_1^x\sigma_1^z\sigma_1^y \nonumber\\
&=1\, ,
\end{align}
which implies that the eigenvalue of $W_{l_c}$ is uniquely determined by the ones of $W_{l_a}$ and $W_{l_b}$. 
A direct consequence of this linear dependence of three loops is that there is no full-flux sector in this model for which all loop operators have eigenvalue -1.
Note that if we would have chosen the alternative definition of the loop operator in Eq.~\eqref{eq:loopOperator2} then we would have concluded that there is no zero-flux sector in this model.
\begin{figure}[t]
\includegraphics[width=\columnwidth]{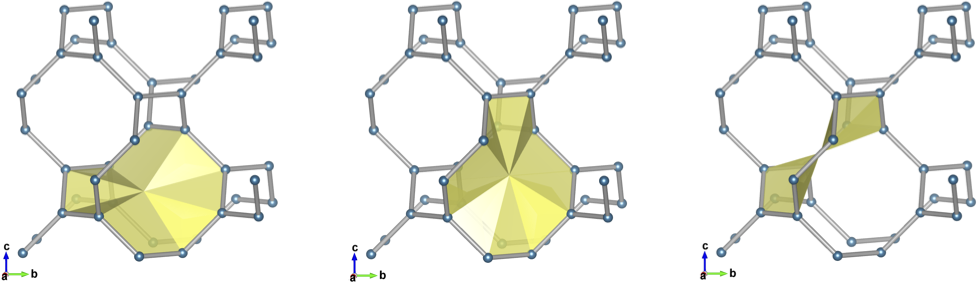}
\caption{(color online) 
	     The minimal closed surface of the hyperoctagon lattice is spanned by three neighboring loops 
	     (with the encompassed volume vanishing).
	     }
\label{fig:Volumes_hyperoctagon}
\end{figure}
This constraint can also be understood graphically. Each loop defines an enclosed surface as illustrated in Fig.~\ref{fig:Volumes_hyperoctagon}. In general, a product of loop operators is constrained if their respective surfaces form a closed `volume'.
For the hyperoctagon lattice, note that the three surfaces corresponding to the respective loops indeed form a closed object though the  
encompassed volume actually vanishes.
This situation should be contrasted to the three-dimensional hyperhoneycomb lattice, which for a complete and self-consistent presentation we discuss in appendix \ref{App:hyperhoneycomb}. In the hyperhoneycomb lattice, four loops of length ten encompass a closed volume as depicted in Fig.~\ref{fig:Loops_hyperhoneycomb} in the appendix and are thus linearly dependent. However, with an even number of linear dependent loops, the corresponding spin model allows for both a zero-flux and a full-flux sector, while in our case only one of the two sectors can exist.

In order to gain a better insight into the physics underlying the different magnetic flux sectors, let us note that the elementary loops of length ten can be uniquely labeled by their midpoint $\mathbf R_l$.
These midpoints form a (deformed) hyper\-kagome lattice. The constraint on the loop operator eigenvalues is then enforced on each of the triangles of the hyperkagome lattice: there are either zero or two loops per triangle that carry flux.  We can thus count the number of independent configurations by noting that there are six midpoints (=loops) and four triangles (=constraints) per unit cell. Thus, there are only $6-4=2$ loop eigenvalues per unit cell that can be chosen freely. As a result there are in total $2^{2N}$ distinct flux sectors, where  $N$ is the number of unit cells. 
In order to determine in which magnetic flux sector the ground state resides, we cannot follow the same route as taken in two spatial dimensions and resort to Lieb's theorem stating that the ground state always resides in the flux-free sector \cite{Lieb}, but instead have
to carefully consider the energetics of the different flux sectors / loops.
For the great majority of points in parameter space, we  numerically observe that creating or enlarging loops costs energy and as such the ground state lies in the flux-free sector also in three spatial dimensions. We therefore restrict our following discussion to the zero-flux sector. 

Let us briefly consider the effective theory for the magnetic flux excitations arising from flipping a loop operator eigenvalue from $+1$ to $-1$.
We note that the midpoints of loops with loop operator eigenvalue $-1$ form themselves closed loop configurations, which live again on the links of a hyperoctagon  lattice -- albeit with opposite chirality to the original one. Due to the constraint, only closed loop configurations are allowed, i.e. there are no magnetic monopoles in this $\mathbb Z_2$ theory.  


\subsection{Majorana representation}

We now proceed to discuss the exact analytic solution of Hamiltonian \eqref{eq:SpinHamRealSpace}.
In analogy to the two-dimensional Kitaev model, such an analytical solution is possible by recasting the 
original spin degrees of freedom in terms of Majorana fermions -- a step that effectively reduces the interacting
spin system to a free fermion problem, which is given by Majorana fermions hopping in a static gauge field~\cite{Tsvelik}.
The fermion system can thus be diagonalized in a straight-forward way thereby also revealing the physics of the interacting spin model.

As first step, we rewrite the original spin degrees of freedom by introducing four Majorana fermion degrees of freedom
$\alpha^x, \alpha^y, \alpha^z$ and $c$ per spin $\sigma$, 
such that 
\begin{align}
\label{eq:4Maj}
\sigma^\alpha=ia^\alpha c, 
\end{align}
where $\alpha=x,y,z$ denotes the spin component. 
The four-dimensional Hilbert space of the four Majorana fermions can be projected back to the two-dimensional physical Hilbert space
of the original spin degrees of freedom by requiring
\begin{align}
\label{eq:D}
  D|\xi\rangle = |\xi\rangle \quad\quad {\rm with} \quad\quad D=a^x a^y a^z c^{\phantom x} .
\end{align}
As we need to introduce four Majorana fermions per site, we have to introduce additional labels to indicate the unit cell index $j$ as well as the unit cell position $\mathbf R$, i.e. $a_j^\alpha(\mathbf R)$. 
The Majorana fermions obey the usual anti-commutation relations 
\begin{align}
\{a_j^\alpha(\mathbf R),a_k^\beta(\mathbf R ')\}&=2\delta_{j,k}\delta_{\alpha,\beta}\delta_{\mathbf R,\mathbf R'}\nonumber\\
\{c_j(\mathbf R), c_k(\mathbf R')\}&=2\delta_{j,k}\delta_{\mathbf R,\mathbf R'}\nonumber\\
\{c_j(\mathbf R),a_j^\alpha(\mathbf R')\}&=0
\end{align}
of a Clifford-algebra. 

\begin{figure}[t]
\includegraphics[width=.8\columnwidth]{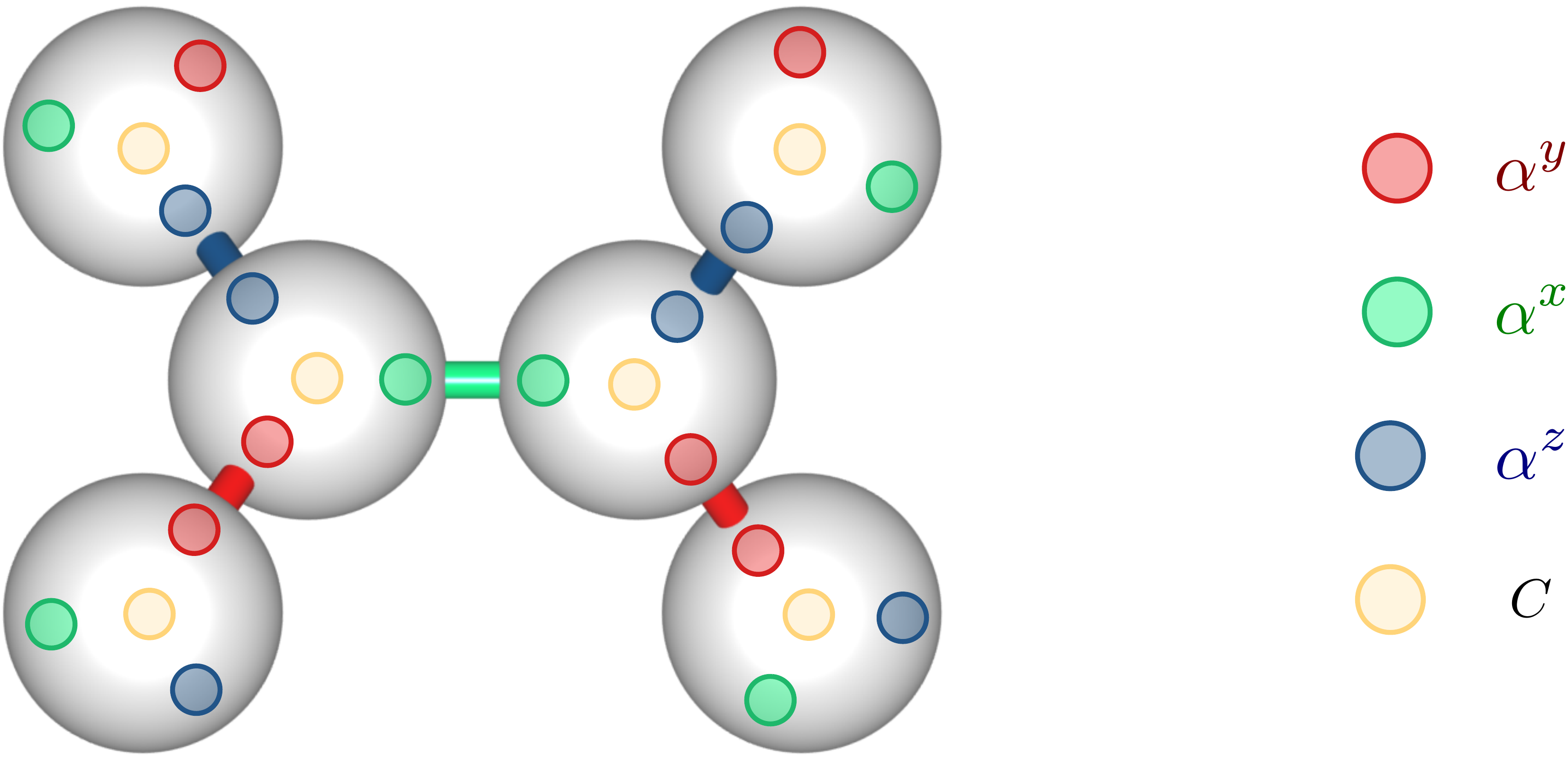}
\caption{(color online) 
	      Illustration of the Majorana fermion representation of the spin degrees of freedom. 
	     }
\label{fig:majoranaRep}
\end{figure}

In terms of the Majorana fermions, the Hamiltonian becomes
\begin{align}
H&=i\sum_{\mathbf{R}} 
J_{x}\left(\hat u_{13} c_{1}(\mathbf{R})c_{3}(\mathbf{R}-\mathbf{a}_{2}) +\hat u_{24} c_{2}(\mathbf{R})c_{4}(\mathbf{R}-\mathbf{a}_{3})\right) \nonumber\\
&+J_{y}\left(\hat u_{12} c_{1}(\mathbf{R})c_{2}(\mathbf{R})+\hat u_{34} c_{3}(\mathbf{R})c_{4}(\mathbf{R}) \right)\nonumber \\
&+J_{z}\left(\hat u_{23} c_{2}(\mathbf{R})c_{3}(\mathbf{R})+ \hat u_{14} c_{1}(\mathbf{R})c_ {4}(\mathbf{R}-\mathbf{a}_{1})\right)\,, 
\label{eq:MajHamRealSpace}
\end{align} 
where we introduced the link operators $\hat u_{ij}=i a_i ^\gamma a_j ^\gamma$ with $\gamma$ being the label of the bond $\langle i,j\rangle$.  The link operators commute among themselves as well as with the Hamiltonian, which implies that we can fix the eigenvalues of all the link operators -- i.e. choose a specific `reference configuration' -- and compute the spectrum of the resulting quadratic Hamiltonian for any given $\{u_{ij}\}$ sector. 
When doing this, one needs to define a direction on the bonds, as $\hat u_{ij}=-\hat u_{ji}$. We choose the convention that the $xx$-bonds are directed along the $\hat y$-direction, the $yy$-bonds along the $\hat z$-direction, and the $zz$-bonds along the $\hat x$-direction. 
This convention ensures that the following discussion remains symmetric in permutations of $J_x$, $J_y$, and $J_z$. 

One may think of the link degrees of freedom as a static $\mathbb Z_2$ gauge field. The gauge transformations are generated by the $D$ operators and Eq. \eqref{eq:D} is equivalent to demanding the physical states to be gauge invariant. In fact, the gauge invariant objects are precisely the loop operators \eqref{eq:loopOperator} introduced earlier. Choosing a reference configuration is equivalent to choosing a specific gauge. The physical properties, such as the Majorana excitation spectrum, are independent of the specific gauge choice, as was already pointed out in Kitaev's original solution~\cite{KitaevModel} of the honeycomb model. 

As we restrict our discussion to the flux-free sector, we may choose all link operators to have eigenvalues $+1$.  Using the  Fourier transformation 
\begin{align}
\label{eq:FT}
c_j(\mathbf R)=\frac 1 {\sqrt{N}} \sum_k e^{i \mathbf k\mathbf R} c_j (\mathbf k)
\end{align}
with $N$ being the number of unit cells,  we can compute the Majorana Hamiltonian in momentum space
\begin{align}
H&=i\sum_{\mathbf{k}} 
J_{x}\left(-e^{-2\pi i k_2} c_{1} (-\mathbf{k})c_{3}(\mathbf{k}) - e^{-2\pi i k_3} c_{2} (-\mathbf{k})c_{4}(\mathbf{k})\right) \nonumber\\
&+J_{y}\left(- c_{1}(-\mathbf{k})c_{2}(\mathbf{k})+ c_{3}(-\mathbf{k})c_{4}(\mathbf{k}) \right)\nonumber \\
&+J_{z}\left( c_{2}(-\mathbf{k})c_{3}(\mathbf{k})- e^{-2\pi i k_1} c_{1}(-\mathbf{k})c_ {4}(\mathbf{k})\right)\,, 
\label{eq:MajHamMomSpace}
\end{align} 
where $k_j$ is defined as the coefficient of the reciprocal lattice vectors $\mathbf k=\sum_{j=1}^3 k_j \mathbf q_j$ with
\begin{align}
\mathbf q_1 &=2\pi (1,-1,0)\,,\nonumber\\
\mathbf q_2 &=2\pi (0,1,-1)\,,\nonumber\\
\mathbf q_3 &=2\pi (0,1,1)\,. 
\end{align} 
Arriving at Hamiltonian \eqref{eq:MajHamMomSpace} has thus reduced the original problem to a four-band Hamiltonian that can
be easily diagonalized \cite{KitaevModel}.


\subsection{Phase diagram}

From the diagonal form of the Hamiltonian \eqref{eq:MajHamMomSpace} we can readily read off the elementary structure of 
the phase diagram in $(J_x, J_y, J_z)$-parameter space by carefully analyzing the excitation spectrum of the Majorana sector. 
In particular, we observe that the Hamiltonian allows for zero-energy solutions indicative of a gapless phase in a range of parameters,
while both the excitations of the Majorana sector and the magnetic flux sector remain gapped in other parts of the phase diagram.

\begin{figure}[t]
\includegraphics[width=\columnwidth]{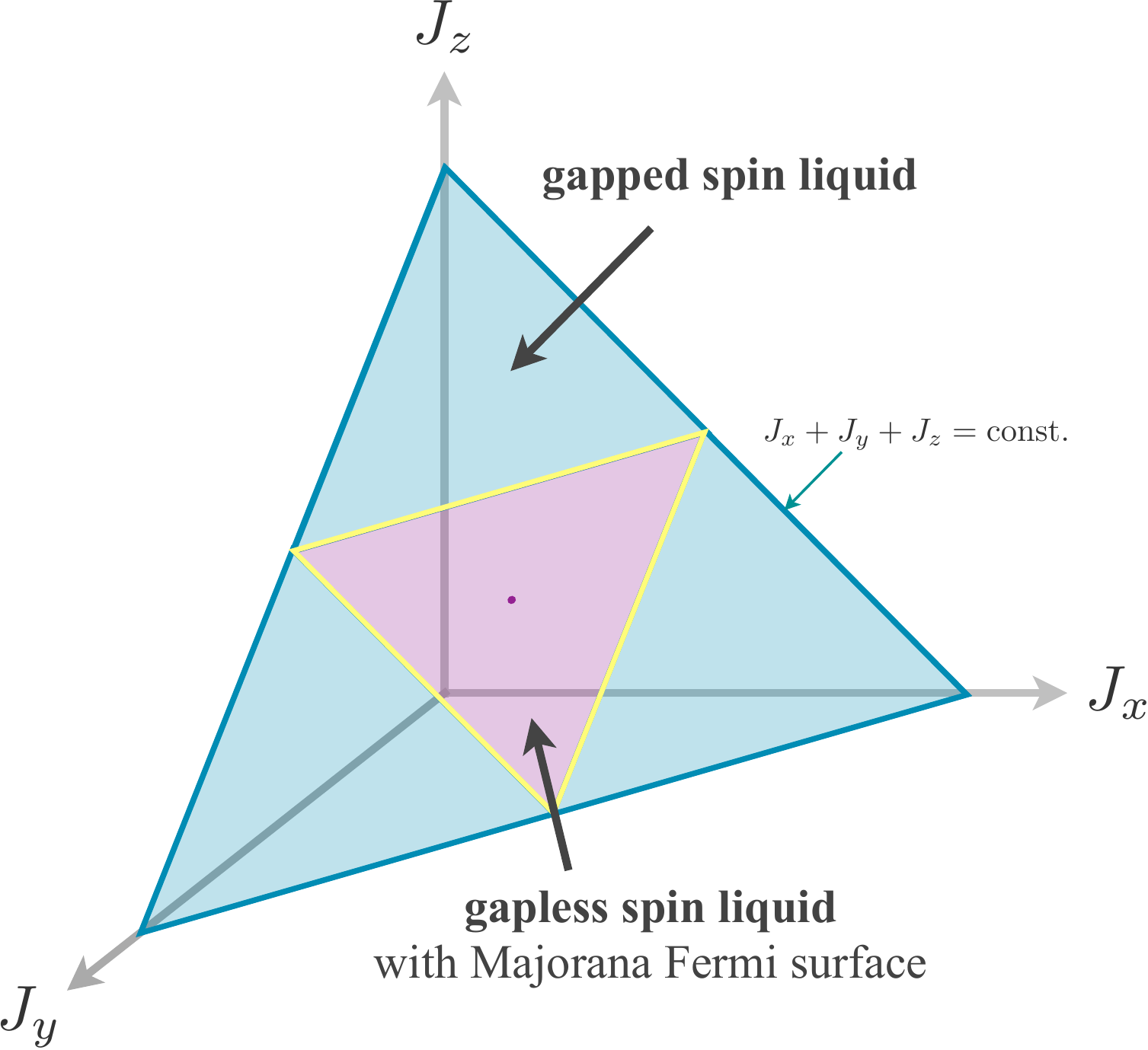}
\caption{(color online) Phase diagram of the Kitaev model on the hyperoctagon lattice.
  	      The gapless spin liquid phase with a Majorana Fermi surface extends in the triangle
	      around the point of isotropic coupling $J_x = J_y = J_z$ (indicated by a point).
	      Around the three corners of the phase diagram where one of the couplings dominates
	      extends a gapped spin liquid phase, which is separated from the gapless phase via a 
	      line of (continuous) phase transitions indicated by the yellow line.
	      }
\label{fig:PhaseDiagram}
\end{figure}

The occurrence of zero-energy solutions is equivalent to requiring that $\det H(\mathbf k)=0$, which becomes 
\begin{align}
\label{eq:zeroE}
\det[ H(\mathbf k)]&=\frac 1{16} (J_x^4+J_y^4+J_z^4+2J_y^2 J_z^2 \cos(k_x) \nonumber\\
& +\, 2J_x^2J_z^2\cos(k_y)+2J_x^2 J_y^2 \cos(k_z) )\nonumber\\ 
&\equiv  0
\end{align}
in cartesian coordinates. In order to determine whether or not the above equation has solutions, let us analyze the limiting behavior of the determinant. Expression \eqref{eq:zeroE} is bounded from above by 
\begin{align}
\label{eq:upper_bound}
\det[ H(\mathbf k)]& \leq \frac 1{16} (J_x^2+J_y^2+J_z^2)^2,
\end{align}
 when setting $\cos(k_x)=\cos(k_y)=\cos(k_z)=1$ and bounded from below by 
 \begin{align}
\label{eq:lower_bound}
\det[ H(\mathbf k)]& \geq - \frac 1{16} (J_x+J_y -J_z)(J_x+J_z-J_y) \nonumber\\
&\times (J_y+J_z-J_x)(J_x+J_y+J_z), 
\end{align}
when setting $\cos(k_x)=\cos(k_y)=\cos(k_z)=-1$. As the upper bound is always strictly positive, there is a zero-energy solution iff the lower bound is negative (or zero).
The latter is equivalent to requiring the triangular inequality
\begin{align}
\label{eq:triangular}
|J_x|+|J_y|&\geq |J_z| \,, \nonumber\\
|J_x|+|J_z|&\geq |J_y| \,, \nonumber\\
|J_y|+|J_z|&\geq |J_x|.
\end{align}
 This is straightforward to derive from \eqref{eq:lower_bound} in case all coupling  constants are positive. For the  case that at least one of the coupling constants is negative, we note that \eqref{eq:zeroE} depends only on the squares of the coupling constants. As a consequence, the upper and lower bound are independent of the signs of the coupling constants.
 
\begin{figure*}[t]
\includegraphics[width=.95\textwidth]{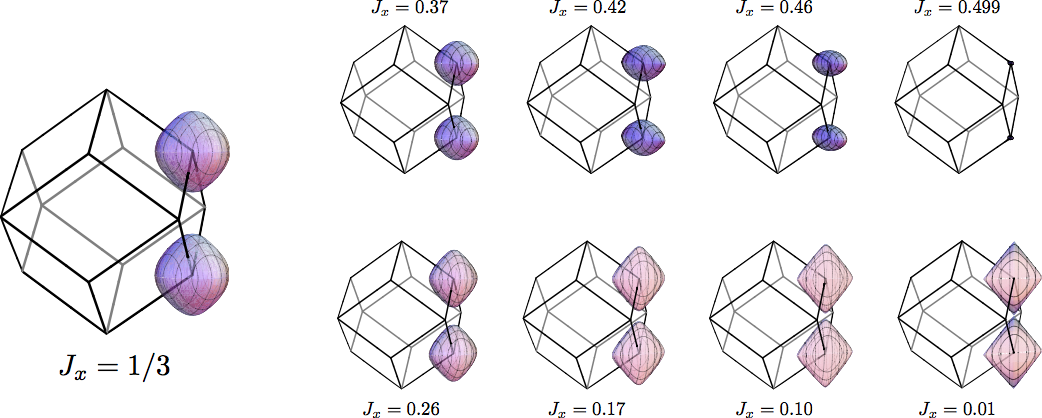}
\caption{(color online) 
               Evolution of the Majorana Fermi surface with varying coupling parameters $J_x$ and $J_y = J_z = (1-J_x)/2$.
               As the phase transition to the gapped spin liquid is approached with increasing $J_z$ (upper row) 
               the Majorana Fermi surface
               shrinks to a single point at momenta $\mathbf Q_1 =( \pi,  \pi,  \pi)$ and $\mathbf Q_2=(\mathbf \pi,\pi,-\pi)$.
               Approaching the decoupling point for $J_x \to 0$ (lower row) the Majorana Fermi surface flattens to two planes.
               }
\label{fig:MajoranaFermiSurface}
\end{figure*}

The triangular inequality \eqref{eq:triangular}  defines the general shape of the phase diagram as depicted in Fig.~\ref{fig:PhaseDiagram}. In the region around the isotropic point $J_x = J_y = J_z$, there are gapless modes in the Majorana sector and the ground state is a gapless spin liquid. If one of the three coupling dominates the Majorana spectrum remains gapped and the ground state is a gapped spin liquid. The gapped and gapless phases are connected via lines of phase transitions which are parametrized by  the equalities in the triangular inequality \eqref{eq:triangular}.

It should be noted that the fundamental shape of this phase diagram is precisely the same one as the ones found for the two-dimensional honeycomb lattice \cite{KitaevModel} and the three-dimensional hyperhoneycomb lattice \cite{Mandal09}. What sets the phase diagrams apart is the actual nature of the two principle gapped and gapless phases for the respective lattices as we will discuss in the following section.
 

\subsection{Gapless spin liquid and Majorana Fermi surfaces}
The fundamental distinction in the phase diagram of the hyperoctagon model is the nature of the gapless phase in the vicinity 
of the isotropic coupling point ($J_x = J_y = J_z$). Its main feature is an extended two-dimensional Majorana Fermi surface of gapless modes.
To see the emergence of such a Fermi surface in the Majorana spectrum around the  isotropic coupling point one needs to invert Eq.~\eqref{eq:zeroE}, which gives 
\begin{multline}
\cos k_z=\\  \left[\frac{J_x^4+J_y^4+J_z^4+2J_y^2 J_z^2 \cos(k_x) +2J_x^2J_z^2\cos(k_y)}{2J_x^2 J_y^2} \right] \,.
\label{eq:surface}
\end{multline}
In combination with the requirement that the lower bound in Eq.~\eqref{eq:lower_bound} becomes negative or zero this parametrizes an entire manifold of $\mathbf k$-points, or more precisely two distinct, non-intersecting continuous surfaces in momentum space centered around the corners of the Brillouine zone at $\mathbf Q_{1/2} = \pi (1,1,\pm 1)$, respectively, as illustrated in Fig.~\ref{fig:MajoranaFermiSurface}.  

It is important to note that while the two surfaces are symmetry related they cannot be mapped onto each other by a reciprocal lattice vector.
As a direct consequence momentum conservation ensures that the zero-energy modes cannot gap out in a pairwise fashion and the surfaces have to remain stable throughout the gapless region. 
Indeed varying the coupling constants away from the point of isotropic coupling only deforms the surfaces, but does not destroy them. 
This is illustrated in the sequence of panels of Fig.~\ref{fig:MajoranaFermiSurface} where we plot the evolution of the surfaces
along a line in parameter space defined by $J_x$ and $J_y=J_z=(1-J_x)/2$. 
Starting from the isotropic point and increasing $J_x$ elongates the surface along the $\hat x$-direction and contracts it in the orthogonal $\hat y$- and $\hat z$-directions as illustrated in the upper panel of Fig.~\ref{fig:MajoranaFermiSurface}. 
Upon further increasing $J_x$ the surface contracts towards the corner of the Brillouine zone. 
As one approaches the phase transition to the gapped spin liquid at $J_x=1/2$, the surfaces have reduced to the points $\mathbf Q_{1/2}$ at the corners of the Brillouine zone. 
On the other hand, decreasing $J_x$ from the isotropic point flattens the surface in the $\hat x$-direction as illustrated in the lower panel of Fig.~\ref{fig:MajoranaFermiSurface}. As $J_x$ goes to zero the two opposite sides of the surface approach each other and touch precisely at the decoupling point $J_x=0$ (and at which the notion of a two-dimensional surface ultimately breaks down as well). 

\begin{figure*}[t]
\includegraphics[width=.95\textwidth]{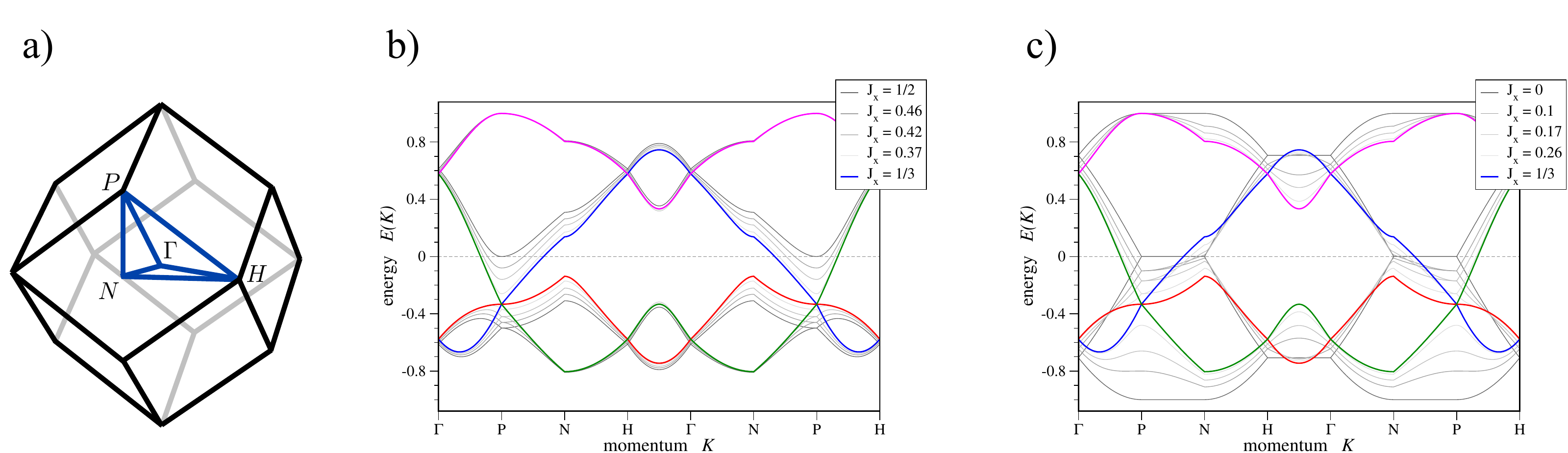}
\caption{(color online) 
	     Band structure of the four principal bands of the Kitaev model along a path 
	      connecting the high-symmetry points $\Gamma=(0,0,0)$, $P= (\pi,\pi,\pi)$, $N=(\pi,\pi,0)$ 
	      and $H= (0,2\pi,0)$ of the Brillouin zone depicted on the left. 
	      The dispersion of the elementary excitations at the Fermi energy is linear for all couplings $0<J_x<1/2$
	      and corresponding couplings $J_y = J_z = (1-J_x)/2$. At the transition to the gapped spin liquid, $J_x=1/2$,
	      the dispersion becomes quadratic.
	      At the point $J_x=0$ the Hamiltonian reduces to decoupled $J_y$-$J_z$ spirals; resulting in a flat spectrum  
	      between the high-symmetry points $P$ and $N$.
	      }
\label{fig:Dispersion}
\end{figure*}

To reveal the nature of the zero-energy gapless modes we plot the dispersion of the four principal bands of Hamiltonian \eqref{eq:MajHamMomSpace}  along certain high-symmetry lines in Fig.~\ref{fig:Dispersion}. 
For the entire gapless phase the dispersion of the bands crossing zero energy is always linear along the direction normal to the surface --
reminiscent of the energy spectrum of a Fermi liquid in the vicinity of the Fermi energy. One should however keep in mind that the four bands
in our Hamiltonian are not spanned by conventional fermionic degrees of freedom, but by Majorana fermions. As such the zero-energy surfaces revealing themselves in the energy spectrum should in fact be thought of as Majorana Fermi surfaces. 

As the phase transition to the gapped phase at $J_x = 1/2$ is approached, the relevant Majorana band moves up in energy and at $J_x=1/2$ no longer crosses the zero-energy level, but merely touches  $E=0$ in a single point with a quadratic dispersion. This scenario is in complete analogy to the phase transitions from the gapless to gapped Majorana phases in both the honeycomb and hyperhoneycomb models.


\section{Instabilities of the Majorana Fermi surface}
\label{sec:FermiSurface}

For conventional Fermi liquids it is well appreciated that the Fermi surface is susceptible to a variety of instabilities, 
the most notable of which is the formation of superconductivity. 
As such two questions immediately arise with regard to the Majorana Fermi surface in our hyperoctagon model -- why is
the Majorana Fermi surface stable in the first place and what are its possible instabilities?

We will address the first question -- the stability of the Majorana Fermi surface -- in the following by showing that it can be
tracked  to the underlying lattice symmetries. We will then devote the remainder of this section to a discussion of
possible instabilities of the Majorana Fermi surface focusing on instabilities arising from a reduction in lattice 
symmetries.

\subsection{Stability of the gapless modes}

To discuss the stability of the Majorana Fermi surface let us first recall some basic facts about Majorana fermions.
An immediate consequence of the Majorana condition $c_j^\dagger (\mathbf R)=c_j(\mathbf R)$ in real space is that
the `creation operator' $c_j^\dagger(\mathbf k)$  in momentum space is defined by $ c_j^\dagger(\mathbf k)=c_j(-\mathbf k)$. 
The latter implies that for every energy state $E(\mathbf k)$ there is a `particle-hole-conjugate' partner at $-\mathbf k$ for which
\[
    	E(-\mathbf k)=-E(\mathbf k) \,.
\]

Quite generally additional energy relations might exist, which depend on the underlying lattice geometry. Of particular importance
in our case is the bipartite nature of the lattice. For a Majorana Hamiltonian on a bipartite lattice with vanishing intra-sublattice hopping amplitudes, one can verify that 
\[    
	E(\mathbf k)=E(-\mathbf k - \mathbf q/2) \,,
\]
where $\mathbf q$ is the reciprocal lattice vector of the translation relating the two sublattices.
Note that for the honeycomb and hyperhoneycomb lattice we have two-site and four-site unit cells, which allow $\mathbf q=0$ and, combining the two energy relations found above, $E(\mathbf k)=-E(\mathbf k)$. This implies that zero-energy modes always occur in pairs at a given momentum. 
On the other hand, it is important to note that for the hyperoctagon lattice $\mathbf q$ generically does not vanish,
because its elementary four-site unit cell is not consistent with a bipartite coloring of the lattice, see also Fig.~\ref{fig:KitaevModel}. For the specific example discussed in the previous section, $\mathbf q/2=(0,0,2\pi)$. As a consequence, the zero-modes are in general all separated in momentum space. 
In order to see what effect this has on the stability of the gapless modes, we start by revising the situation for the honeycomb model following Kitaev's original arguments and afterwards extend this discussion to the three-dimensional generalizations. 

\subsubsection*{Stability of the gapless modes in the honeycomb model}
In the honeycomb model, the Majorana Hamiltonian in momentum space is a $2\times2$ matrix of the form
\begin{align}
\label{eq:honeycomb}
H&=\left(\begin{array}{cc} \mathbf{0}& i f(\mathbf k) \\ -i f^\star (\mathbf k)  &\mathbf{0} \end{array}\right), 
\end{align}
where $f(\mathbf k)$ is a complex-valued function. The vanishing diagonal elements are in fact protected by time-reversal symmetry \cite{KitaevModel}. The eigenvalues of the Hamiltonian are given by $E(\mathbf k)=\pm |f(\mathbf k)|$ and zero-energy modes occur when $f(\mathbf k)=0$. 

Let us start by noting that the conditions $\mbox{Re}[f(\mathbf k)]=0$ and $\mbox{Im}[f(\mathbf k)]=0$ define (several) closed lines in momentum space, denoted in the following by $\Gamma_R$ and $\Gamma_I$. The zeroes of $f(\mathbf k)$ are then, in general, given by the intersections of these lines. As a consequence, each pair of zero-modes at momentum $k$ comes with a partner, which in fact is located at $-\mathbf k$. Changing parameters deforms the line $\Gamma_{R/I}$, which in turn moves the zeroes. The only way to gap out the system is by moving the lines $\Gamma_{R/I}$ sufficiently, such that they do not intersect any longer.  The phase transition corresponds to a situation, where $\Gamma_R$ and $\Gamma_I$ merely touch. 
This structure of the eigenenergies readily confirms that  \emph{separated pairs} of zero-modes are stable in  a finite parameter regime. 
On the other hand, a similar line of reasoning shows that \emph{lines} of zero-modes are not stable and can generically be gapped out completely by even an infinitesimal change in parameters. 
There is, however, a generic way to stabilize lines of gapless modes in two-dimensional generalizations of the Kitaev model, which we will comment on below. 

\subsubsection*{Stability of the gapless modes in the hyperhoneycomb model}
We now extend the above discussion to the three-dimensional models, first considering the hyperhoneycomb model. In fact, the arguments of this section are valid for any three-dimensional Kitaev-type model on a bipartite lattice, which is time-reversal invariant and  where the unit cell is compatible with a bipartite coloring of the lattice. These assumptions are sufficient to determine the Majorana Hamiltonian to be a block matrix, where only the off-diagonal matrices are non-vanishing
\begin{align}
\label{eq:bipartiteTRIham}
H&=\left(\begin{array}{cc} \mathbf{0}& \mathbf{A} \\ \mathbf{A}^\dagger &\mathbf{0} \end{array}\right).
\end{align}
where $A$ is a complex matrix. The eigenvalues of the Hamiltonian are given by $E(\mathbf k)=\pm |\lambda_j(\mathbf k)|$, where $\lambda_j(\mathbf k)$ are the eigenvalues of $A$. Analogously to the two-dimensional case, there are zero-energy solutions for
$\lambda_j(\mathbf k)=0$. However, as the model is three dimensional the conditions Re$[\lambda_j(\mathbf k)]=0$ and Im$\lambda_j(\mathbf k)=0$ now define (several) surfaces in momentum space, denoted again by $\Gamma_{R/I}$. Zero-energy modes occur at the intersection of $\Gamma_R$ with $\Gamma_I$ and thus in general form closed lines in momentum space. Changing parameters in the model deforms the surfaces $\Gamma_{R/I}$ and, thus, the corresponding line of gapless modes, but cannot induce a gap in the system. The latter can only be done by changing parameters sufficiently, such that the surfaces $\Gamma_R$ with $\Gamma_I$ do not intersect any longer,  during which the line shrinks to a single point and vanishes. This shows that extended lines of gapless modes are topologically stable in these types of models. A similar line of reasoning substantiates that  this does not apply to separated points as well as surfaces of gapless modes -- both of which are not stable objects for these kinds of models and can only be accidental. 

\subsubsection*{Stability of the gapless modes in the hyperoctagon model}
The important distinction of the hyperoctagon model to the ones discussed above, is that the unit cell is not compatible with a bipartite coloring of the lattice. Time-reversal invariance thus only requires the diagonal elements to vanish. The Hamiltonian can then be written as 
\begin{align}
\label{eq:honeyoctagon}
H&=\left(\begin{array}{ccc} 0 && \mathbf A\\ & \ddots & \\ \mathbf A^\dagger &&0  \end{array}\right), 
\end{align}
which is a band Hamiltonian with the additional property $E(\mathbf k)=-E(-\mathbf k)$ due to the Majorana condition \cite{FootnoteZeroes}. Zero-energy modes occur, when bands cross the $E=0$ line, which results in {\em surfaces} of gapless modes. In general, the incompatibility of the unit cell and the bipartite coloring of the lattice implies that $E(\mathbf k)\neq-E(\mathbf k)$. As a result, there is generically only a \emph{single} Majorana zero-mode at a given momentum. The surfaces are thus trivially stable -- changing parameters deforms the energy bands and thus the surfaces, but gapping a surface can only be done by either shrinking it to a point or by superimposing two such surfaces, in which case there are two gapless Majorana modes at the same momentum. 
The latter is not stable and can always be gapped out by an infinitesimal change in parameters. Likewise, one can verify that lines or separated points of gapless modes are not stable objects for these types of Majorana Hamiltonians. 

This line of reasoning also sheds light on how to obtain two-dimensional models with a stable Fermi surface. In analogy to the case above, one needs to consider two-dimensional lattices, where the unit cell is not compatible with the bipartite coloring of the lattice. 
An example would be the square-octagon lattice studied in \cite{SquareOctagonModel}. Considering various flux sectors, which do not enlarge the unit cell, indeed demonstrates that separated points of gapless modes are not stable, while closed lines are.

\subsection{Fermi surface instabilities due to unit cell doubling}

Following the line of reasoning in the previous subsection points to a natural way to destabilize the Fermi surface by enlarging the unit cell such that the two Majorana Fermi surfaces are mapped onto each other. 
This requirement is identical to identifying an enlarged unit cell that allows for a bipartite coloring of the lattice within that unit cell (and thus $\mathbf q$ vanishes) and the disappearance of a lattice symmetry that prohibits the zero-energy modes from gapping out pairwise.
Below we will see that the Majorana Fermi surface is indeed no longer stable when enlarging the unit cell. However, the surface does not gap out completely, but instead reduces to a line of gapless modes -- similar to the situation of the Kitaev model on the hyperhoneycomb lattice. A closed line of gapless modes is, on the other hand, a topologically stable object for three-dimensional Hamiltonians such as \eqref{eq:bipartiteTRIham}. Changing parameters can only deform the line, but not gap it out. 

\begin{figure}[t]
\includegraphics[width=\columnwidth]{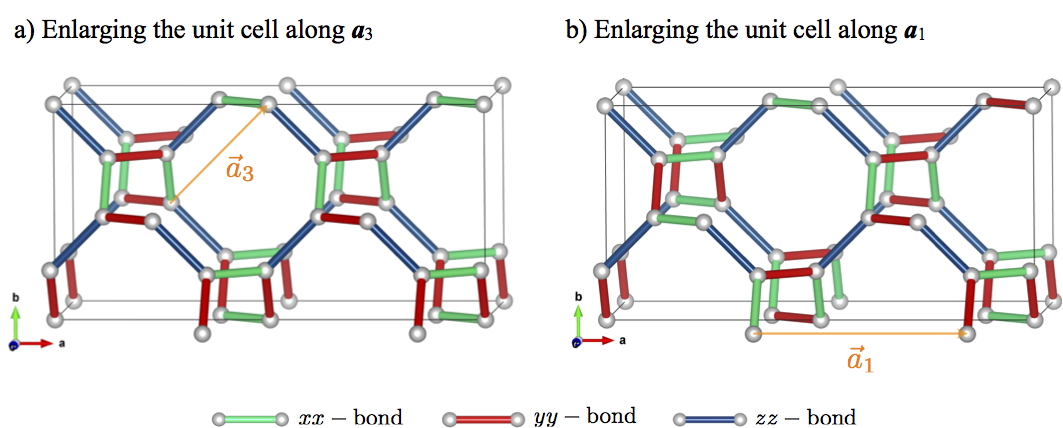}
\caption{(color online) Illustration of the different couplings in the Kitaev model, when enlarging the unit cell along the a) $\mathbf a_3$ and b) $\mathbf a_1$ direction. The arrow marks the broken translation vector. The unit cell of the model in a) is compatible with a bipartite coloring of the lattice while the one in b) is not. }
\label{fig:Kitaev2}
\end{figure}

In order to elucidate this, let us consider an alternative covering of the hyperoctagon lattice with $xx$, $yy$, and $zz$ bonds, which is no longer  invariant under $\mathbf a_3=\frac 1 2 (1,1,1)$ translations. The enlarged unit cell thus allows for a bipartite coloring of the lattice.  A specific realization of this is shown in Figure \ref{fig:Kitaev2} a). Compared to the original model discussed in Section \ref{sec:the_model},  the $xx$ and $yy$ bonds are switched in every second unit cell along the $\mathbf a_3$ direction.  This enlarges the unit cell with the new translation vectors becoming $\mathbf a_1=(1,0,0)$, $\mathbf a_2=(0,1,0)$ and $\mathbf a_3 =(0,0,1)$. Note that these are the simple cubic translations, i.e. we have moved from a body-centered cubic structure to a simple cubic one by doubling the unit cell along the $\mathbf a_3$ direction.

The model still fulfills that no vertex is connected to the same bond type twice and is, therefore, still exactly solvable with the methods used above. 
In contrast to the original Hamiltonian \eqref{eq:SpinHamRealSpace}, the new Hamiltonian is no longer  isotropic in the coupling constants, as can already be deduced from Fig.~\ref{fig:Kitaev2} a) -- the resulting Kitaev model is only symmetric in $J_x\leftrightarrow J_y$, while the $zz$ type bond stands out. Thus, we expect qualitatively different behavior when $J_x=J_y$ compared to $J_x\neq J_y$. 

Let us first comment on the flux sectors on the model. It can be shown that the constraints on the loop operators are independent on the choice of the covering of the hyperoctagon lattice in $xx$, $yy$, and $zz$ bonds, even though the loop operators themselves change. In analogy to our original discussion in Section \ref{sec:the_model} we restrict ourselves to the vortex-free sector in the following and focus on the changes in the Majorana sector induced by the alternation of the Kitaev interactions. In particular, we are interested in the gapless modes in the Majorana sector. 

\begin{figure}[t]
\includegraphics[width=\columnwidth]{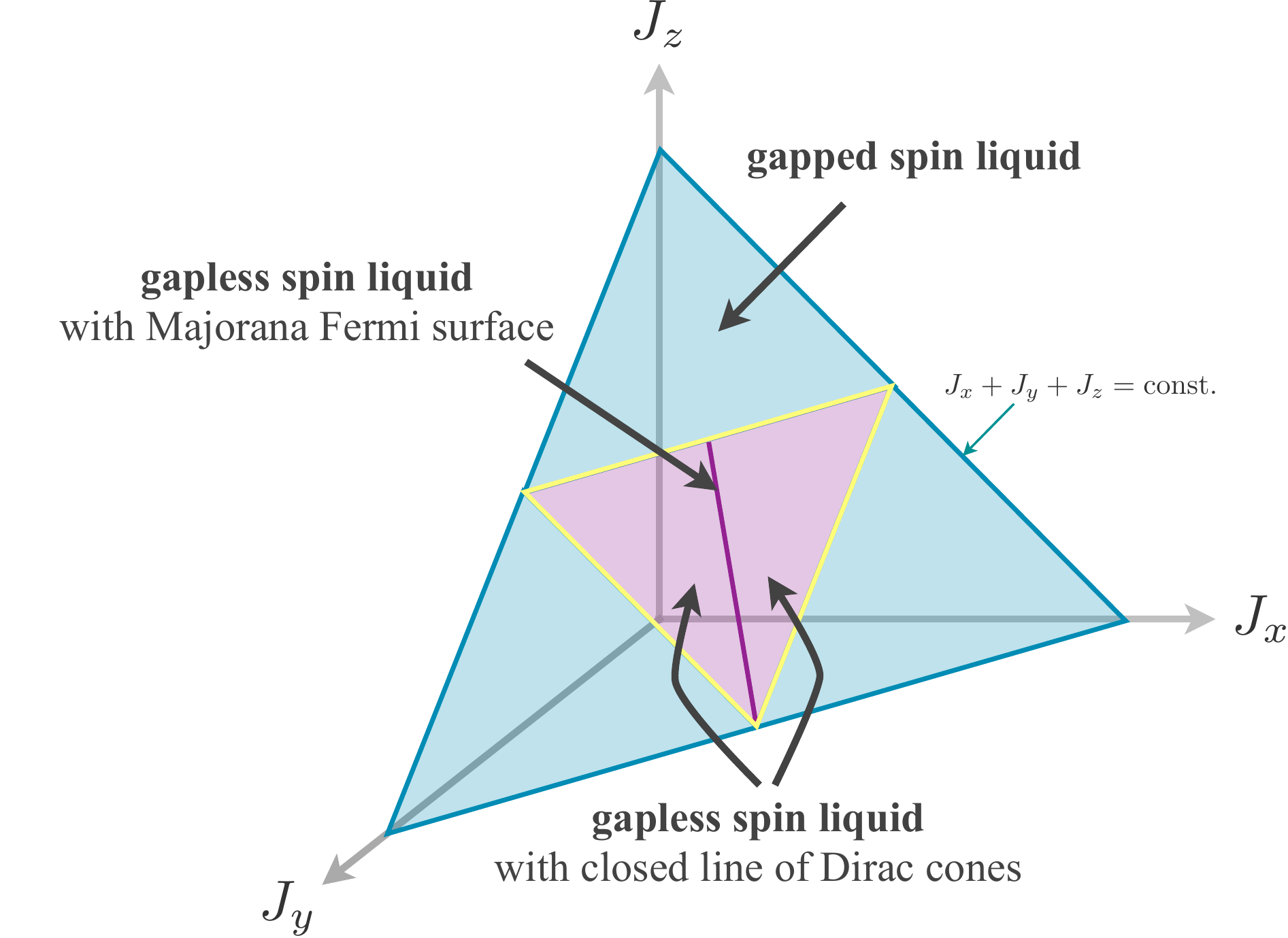}
\caption{(color online) 
Sketch of the phase diagram of the model defined in Fig. \ref{fig:Kitaev2} a). Along the line $J_x=J_y$, there is a Majorana Fermi surface, which is identical to the original model defined in Section \ref{sec:the_model}. Away from this line, the Majorana surface is partly gapped out and reduced to a closed line of Dirac cones. 
}
\label{fig:PhaseDiagramLine}
\end{figure}

The parameter region, in which gapless modes exist, is identical to the original model, i.e. determined by the triangular inequality \eqref{eq:triangular}. The transition lines to the gapped phases are parametrized by the equalities of Eq. \eqref{eq:triangular}. 
In addition, the behavior of the model along the $J_x=J_y$ line in parameter space is identical to the original model. The latter can be understood by noting that the modified Majorana Hamiltonian at this line `looks' like the original one, albeit with an enlarged unit cell. In particular, on the line $J_x=J_y$ we still find a surface of gapless Majorana modes when $J_x=J_y$. However, the surface is not stable against departing from that parametrization condition.  Even an infinitesimal discrepancy in the coupling constants $J_x\neq J_y$ immediately opens up a gap on most of the surface and the two-dimensional Majorana Fermi surface collapses onto a line of gapless excitations, with linear dispersion in the normal directions. This line of gapless modes then remains stable throughout the rest of the gapless phase and contracts to a point on the transition lines to the gapped phases. The phase diagram for this model is visualized in  Fig.~\ref{fig:PhaseDiagramLine}. 

\begin{figure}[t]
\includegraphics[width=\columnwidth]{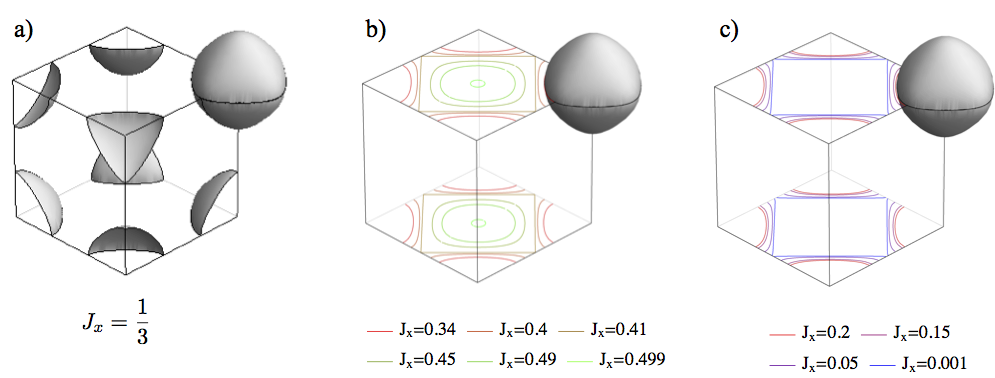}
\caption{(color online) 
Visualization of the surface vs. lines of gapless modes. On the line $J_x=J_y$, the model exhibits a Majorana Fermi surface -- centered around $(\pi,\pi,\pi)$  as shown in a). Away from $J_x=J_y$, the surface reduces to a line, which lies in the $k_z=\pm \pi$ plane. Panels b) and c) show the behavior of the gapless line, when b) increasing $J_x$ and c) decreasing $J_x$, and setting $J_y=J_z=(1-J_x)/2$. 
}
\label{fig:Enlarged_gapless_lines}
\end{figure}

This behavior should be contrasted to what happens when increasing the unit cell in the $\mathbf a_1$ direction, such that the enlarged unit cell is still incompatible with a bipartite coloring of the lattice. One possible way to achieve this is to switch the $xx$ and $yy$ bonds in every second unit cell along the $\mathbf{a}_1$ direction, shown in Fig.~\ref{fig:Kitaev2} b). Note that this again breaks the isotropy of the original model in the coupling constants. Similarly to the two previously discussed models, the region of parameter space with gapless excitations is again defined by the triangular inequality \eqref{eq:triangular}. For the same reasons as stated above, the model with the enlarged unit cell has the same properties as the original one along the $J_x=J_y$ line in parameter space. In contrast to the model with enlarged unit cell along the $\mathbf{a}_3$ direction, discussed above, the surface remains stable, even when departing from this parametric condition. The behavior of the surface is similar to the original model. In particular, the dispersion around the gapless surface is linear in the normal direction, except at the phase transition to the gapped phases, where the surface shrinks to a single point with quadratic dispersion. The reason for the stability of the surfaces can again be tracked to their relative displacement in momentum space -- a direct consequence that the unit cell is not compatible with the bipartite coloring of the lattice. 


\section{Discussion and outlook}
\label{sec:outlook}

To a certain extent one can take the perspective that our results for the Kitaev model on the hyperoctagon lattice complete a family of analytically tractable spin liquids of growing complexity
-- starting from a two-dimensional Dirac spin liquid on the honeycomb lattice over the intermediate step of a three-dimensional spin liquid with a line of gapless modes on the hyperhoneycomb lattice to finally a spin liquid with a full, two-dimensional surface of gapless modes for the hyper\-octagon lattice.
Despite this ascent in complexity, we want to point out that all three instances share certain features such as rapidly decaying dynamical spin-spin correlation functions while differing in other aspects such as the nature of the phases induced by time-reversal symmetry breaking perturbations such as a magnetic field.

Starting with the similarities, it is interesting to observe that independent of the nature of the manifold of their gapless modes all three spin liquids exhibit dynamical two-spin correlation functions that are identically zero beyond nearest neighbor separation. This extremely rapid decrease was first observed in the context of the honeycomb model \cite{SpinDynamics}, but the very same arguments employed there also hold for the hyperhoneycomb and hyperoctagon lattices.

Another parallel arises when distorting the couplings in the Kitaev model such that one of the coupling constants becomes dominant
(i.e. such that the triangular inequality does not hold any longer) and the Majorana sector is gapped out, see the phase diagram in 
Fig.~\ref{fig:PhaseDiagram} for comparison. In these gapped phases the low-lying excitations are instead given by configurations, where some of the loop operators \eqref{eq:loopOperator} have negative eigenvalue. While for the two-dimensional honeycomb lattices two types of excitations can be discerned and identified with the point-like (electric and magnetic) excitations of a $\mathbb Z_2$ gauge theory, the effective low-energy theory for the three-dimensional lattice is somewhat more elaborate. Mandal and Surendran argued \cite{Mandal11} that there is only a single loop-like excitation in the gapped phases of the hyperhoneycomb model that exhibits non-trivial (semionic) braiding properties. Likely, a variant of their argument with similar conclusions can be applied to the hyperoctagon model as well. 

A clear distinction between the three spin liquid phases arises when considering the effect of time-reversal symmetry breaking perturbations 
such as a magnetic field. For the honeycomb model such a perturbation is of utmost interest as the gapless Majorana modes are protected by time-reversal and even an infinitesimal magnetic field  (applied along the 111-direction)  gaps out the spin liquid into a gapped topological phase with non-Abelian vortex excitations. 
On a more technical level the reason for this drastic change induced by the magnetic field can be seen in terms of the symmetry class classification \cite{TopoClassification,TopoClassification2} of the underlying free (Majorana) fermion problem \cite{Zirnbauer,AltlandZirnbauer}.
In the absence of time-reversal symmetry breaking the model is in symmetry class BDI, while in the presence of a magnetic field the symmetry class changes to class D. The latter allows for a $\mathbb Z$ classification of topological phases in two spatial dimensions. Indeed the two bands split by the magnetic field are characterized by Chern numbers  $\pm 1$ indicating the non-Abelian nature of the gapped phase.

When considering a similar line of arguments for the three-dimensional models a different picture emerges.  As noted earlier, the zero-energy modes of the hyperhoneycomb model are protected by time-reversal symmetry, while the ones of the hyperoctagon lattice are not. So while we expect the gapless phase of the hyperhoneycomb model to gap out immediately in the presence of a magnetic field, this is far from obvious for the zero-energy modes of the hyperoctagon model as the spectrum is robust against any two-fermion term that does not break translation symmetry. 
Independent of whether a gap opens in the spectrum, it still holds that the symmetry class of the underlying free (Majorana) fermion model changes from class BDI to class D in the presence of a time-reversal symmetry breaking term. However, in contrast to its two-dimensional counterpart symmetry class D does not harbor any topological phases in three dimensions \cite{TopoClassification,TopoClassification2}. 
As a consequence, the three-dimensional systems cannot be driven into a (non-Abelian) topological phase by applying a magnetic field (or any other time-reversal symmetry breaking perturbation). However, one might still be able to employ similar ideas to the ones used by Ryu in Ref.~\cite{Ryu} to stabilize a non-trivial topological phase by introducing additional (orbital) degrees of freedom such that the augmented model can be reformulated as  a free fermion model in symmetry class DIII.  The latter does have a $\mathbb Z$ classification in three dimensions and, thus, allows for three dimensional analogs of the topological phase in the honeycomb model.

\subsubsection*{Thermodynamic signatures of the spin liquid}

Finally, an interesting perspective emerges when recasting our results in the terminology conventionally used to characterize various spin liquid states \cite{SpinLiquids}. In this language, we have discovered a spin liquid with a {\em spinon Fermi surface} that covers an extensive two-dimensional manifold in momentum space. The quest to identify magnetic systems harboring such spinon Fermi surfaces has typically inspired theorists to consider a slave-fermion approach where the fermion interacts with a {\em fluctuating} $U(1)$ gauge field -- a situation that is notoriously hard to track analytically and any progress coming at the expense of compromises on the level of various decoupling/mean-field approaches. 
This situation should be contrasted to the current situation where we have stumbled upon a system with a spinon Fermion surface with a much simpler and analytically exact description in terms of Majorana fermions interacting with a {\em static} $\mathbb Z_2$ gauge field. However, it is important to note that this difference is not a mere conceptual one, but one that has direct implications for thermodynamic observables such as the specific heat coefficient $C/T$.
For a $U(1)$ spin liquid the specific heat diverges as 
\[
   C(T) \propto T \ln(1/T) \,,
\]
i.e. the specific heat coefficient $\gamma = C/T$ {\em diverges} logarithmically at low temperatures \cite{SpecificHeat-U1-SpinLiquid}.
For our case of a spinon Fermi surface emerging from Majorana fermions interacting with a $\mathbb Z_2$ gauge field we find 
\[ 
   C(T) \propto T \,,
\]   
i.e. the specific heat coefficient  $\gamma$ goes to a {\em constant} at low temperatures.
Finally, this situation should be contrasted to the spin liquid with a Fermi line, as it was found for the hyperhoneycomb lattice, where the specific heat grows as \cite{hyperKim}
\[
     C(T) \propto T^2 \,,
\] 
i.e. the specific heat coefficient $\gamma$ {\em vanishes} in the limit of $T\to 0$.
Remarkably enough, this implies that a simple thermodynamic experiment could immediately distinguish these three seemingly equally exotic spin liquids.

Returning to the perspective of the free (Majorana) fermion system underlying our gapless spin liquid, there is one obvious bouquet of questions that we have not addressed in the manuscript at hand -- namely the various pairing instabilities that the Fermi surface of our system might exhibit. One might be particularly interested in asking what instabilities can be induced by additional interactions such as a Heisenberg exchange argued to accompany the Kitaev interactions in any microscopic description of Iridate compounds \cite{Jackeli09,Chaloupka10}.
The effective description of the hyperoctagon model in terms of {\em spinless} fermions suggests $p$-wave pairing as the natural candidate for opening a gap. Interaction terms of this type can indeed arise in a perturbative analysis of the Heisenberg exchange. 
The question of whether or not these terms lead to a collapse of the Majorana Fermi surface or even gap out all Majorana modes in 
the system is left for future work. \\


\noindent{\bf Acknowledgements.--}
We thank A. Akhmerov, A. Altland, P. Becker-Bohaty, S. Parameswaran, F. Pollmann, and especially A. Rosch for insightful discussions. 
ST acknowledges the  hospitality of the Aspen Center for Physics where some of the ideas underlying this 
manuscript were perceived. ST is indebted to L. Balents for a beginner's guide to 3D printing.
We acknowledge partial support from SFB TR 12 of the DFG. 

%
%


\appendix


\section{Possible magnetic materials in space group I4$_1$32}
\label{sec:AppSpaceGroupMaterials}
In this appendix, we want to expand our discussion of possible magnetic materials candidates in space group I4$_1$32 (no. 214).
The guiding idea in our analysis is to put the space group symmetries of I4$_1$32 to work and look for various possible ways to fill the interstitial sites between the network of edge-sharing IrO$_6$ cages as illustrated in Fig.~\ref{fig:O6cages} of the main text. In addition,
we have to take into account that the fundamental building blocks of IrO$_3$ have valency $-2$ and as such are looking for interstitial fillings that allow to chemically compensate this valence.

\subsection{Space group symmetries}

Let us first consider the effect of the space group symmetries. There are in total 48 symmetry operations in the space group I4$_1$32. A generic point in the cubic cell is, thus, mapped to in total 48 distinct points; the set of these points will, in the following, often be labeled by a single representative. However, there are several high-symmetry points, respectively lines in the cubic cell, which are mapped to far fewer points. In total, we can distinguish 5 types of lattice point, according to the number of distinct lattice points that can be reached by the symmetry  operations. 

i) The lattice point is mapped to 8 distinct lattice points in the unit cell. This is possible in two inequivalent ways. The resulting points form the two chiralities of the hyperoctagon lattice. The representatives of the two possibilities are $\frac 1 8 (1,1,1)$ and $-\frac 1 8 (1,1,1)$. In the main text, we chose the hyperoctagon lattice generated by $\frac 1 8(1,1,1)$; thus, in the following we place the Iridium atoms on this set of sites. 

ii) The lattice point is mapped to 12 distinct lattice points in the unit cell. This is again possible in two inequivalent ways. The representatives are given by $(0,\frac 1 4, \frac 1 8 )$ and $(0,\frac 1 4 ,\frac 5 8)$. These sites form effective lattices, which are  deformations of the two chiral versions of the hyperkagome. In fact, they can be identified as the medial lattices of the two chiral hyperoctagon lattices in i). 

iii) The lattice point is mapped to 16 distinct lattice points in the unit cell. There are infinitely many such possibilities, as long as the representative is chosen on the line $(x,x,x)$ (except the high symmetry points already listed in i)). 

iv) The lattice point is mapped to 24 distinct lattice points in the unit cell. There are many high-symmetry lines in the cubic unit cell, which lead to this behavior. The oxygen sites are an example for this type of lattice points -- represented by $\frac 1 8 (1, -1, 1)$. 

v) The lattice point is mapped to 48 distinct lattice points in the unit cell, which applies to all lattice points that don't lie on one of the above mentioned high-symmetry lines.

\subsection{Possible chemical compositions}

\begin{figure}[t]
\includegraphics[width=\columnwidth]{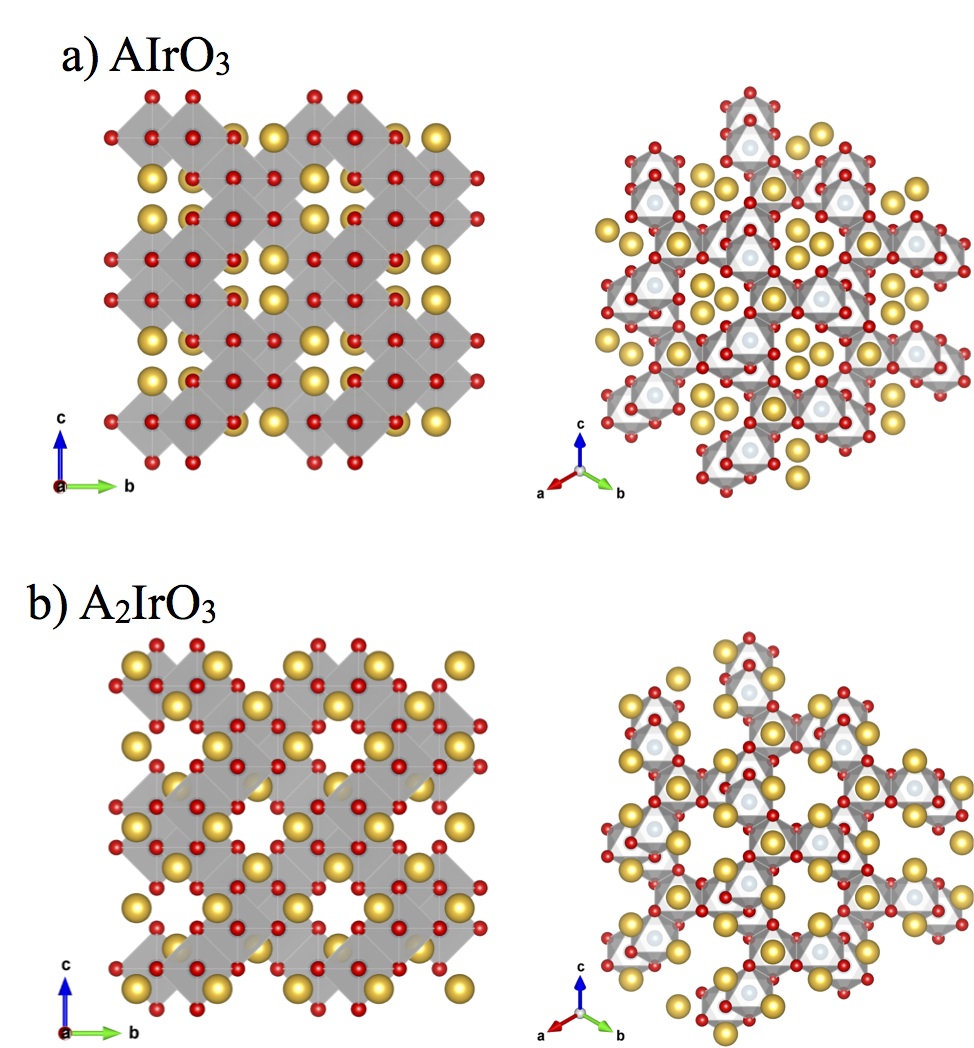}
\caption{(color online) Two possibilities of placing atoms on the interstitial sites in the IrO$_3$ structures, indicated by the grey octahedra. 
Panel a) shows  the crystal structure by placing  atoms of valency $+2$ on the sites of type i) (see text). In panel b), atoms with valency $+1$ are placed on sites of type iii), which are generated by the representative $(0,0,0)$. }
\label{fig:space_group}
\end{figure}

This structure, imposed by the symmetry group, severely restricts the composition of possible compounds. Assuming the presence of edge-sharing IrO$_6$ cages, we note that the above analysis implies that there are 8 IrO$_3$ in a cubic unit cell. Thus, the remaining atoms must compensate a total valency of $-16$. The latter can, for instance, be achieve by placing 8 atoms with valency $+2$ on the remaining set of sites of type i).  The resulting compound is of the form AIrO$_3$, where  A is one of the alkaline-earth elements Ca, Sr or Ba. 
Another  possibility is to place 16 atoms of valency $+1$ on the sites of type iii). This results in a material of the type A$_2$IrO$_3$, where A is one of the alkali atoms Na or Li. The resulting crystal structures for both possibilities are visualized in Fig.~\ref{fig:space_group}.


\section{The Kitaev model on the hyperhoneycomb lattice}
\label{App:hyperhoneycomb}
To complement our discussion of the Kitaev model on the hyperoctagon lattice, we will present a brief, self-contained summary of the
Kitaev model on the hyperhoneycomb lattice, an alternative trivalent 3D lattice, in this appendix. For clarity, we will use similar notations and conventions as in the main text,
but emphasize that in doing so we also closely follow the original solution of the Kitaev model on the hyperhoneycomb lattice as discussed
in some detail in Ref.~\onlinecite{Mandal09}.

\subsection{The hyperhoneycomb lattice}

\begin{figure}[b]
\includegraphics[width=\columnwidth]{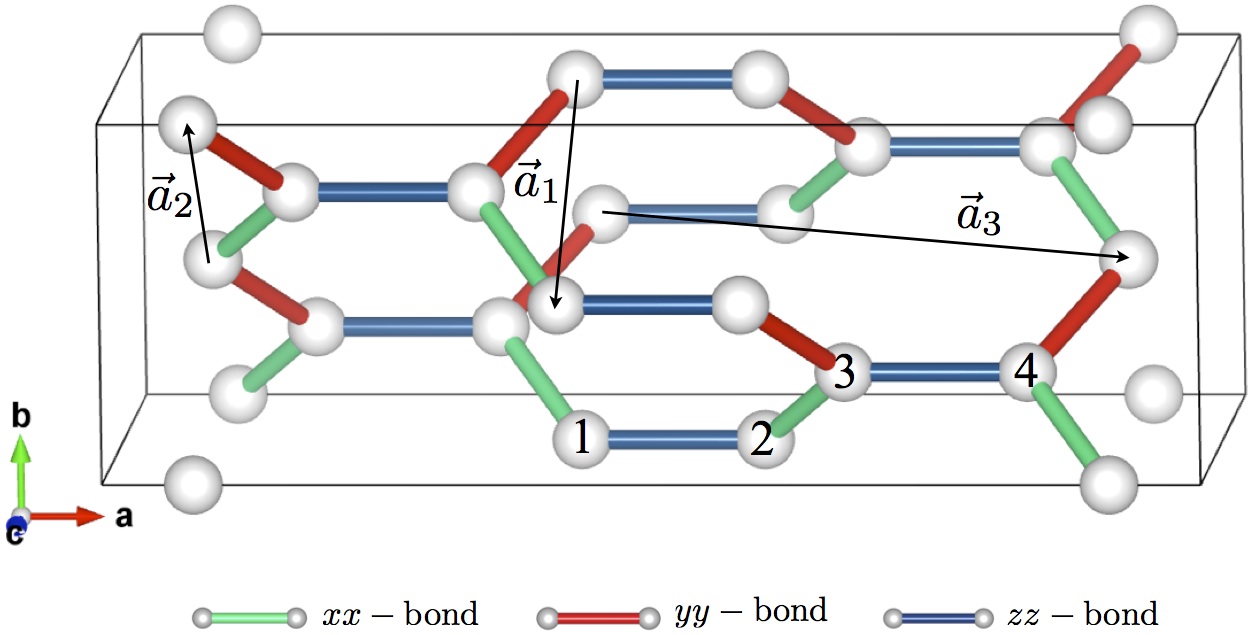}
\caption{(color online) The hyperhoneycomb lattice. The four-site unit cell and translation vectors are indicated.        
Green bonds correspond to $xx$-couplings, red bonds to $yy$-coupling, and blue bonds to $zz$-coupling, respectively.  }
\label{fig:hyperhoneycomb_UnitCell}
\end{figure}

The hyperhoneycomb lattice is an alternative 3D lattice with a trivalent lattice structure which is illustrated in Fig.~\ref{fig:hyperhoneycomb_UnitCell}. Its elementary building blocks are zig-zag chains running along the crystallographic $\mathbf b$- and
$\mathbf c$-axises, which are coupled along the $\mathbf a$-axis. Its crystal structure can be classified as a face centered orthorhombic lattice of space group number 70. Of particular importance for our discussion is the fact that it is not spatially isotropic as the hyperoctagon lattice, but has one `preferred' direction -- the $\mathbf a$ direction in Fig.~\ref{fig:hyperhoneycomb_UnitCell}.  

\begin{figure} 
\includegraphics[width=\columnwidth]{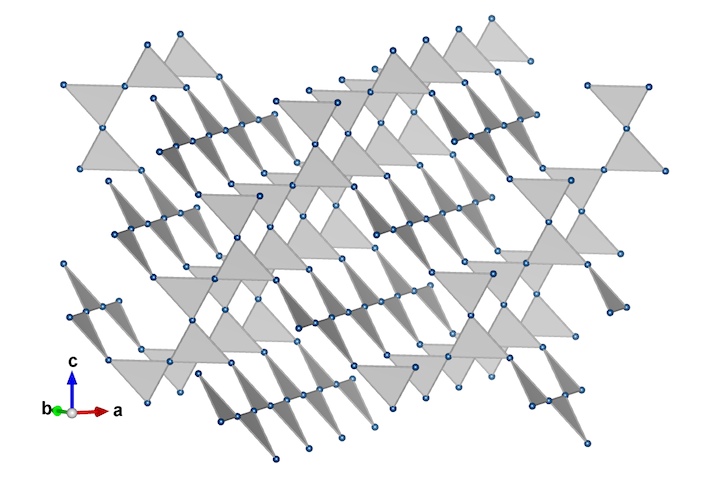}
\caption{(color online) The medial lattice of the hyperhoneycomb lattice, which we dub orthorhombic kagome lattice. The lines of dark-shaded triangles run along the (1,1,0)-direction, the light-shaded ones along the (0,1,-1) direction. The lattice has been slightly deformed for better illustration.}
\label{fig:3DKagome}
\end{figure}

\begin{figure} 
\includegraphics[width=\columnwidth]{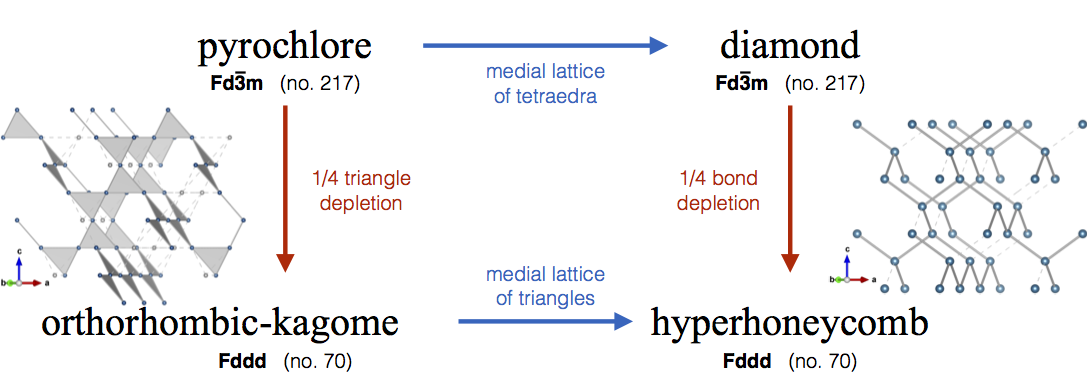}
\caption{(color online) Another set of relations between various three-dimensional lattices -- see also Fig.~\ref{fig:LatticeSummary} in comparison. The hyperhoneycomb lattice is the medial lattice of the so-called orthorhombic-kagome lattice depicted in Fig.~\ref{fig:3DKagome}. Like the hyper-kagome lattice the orthorhombic-kagome lattice can be obtained from the pyrochlore lattice via depletion of 1/4 of the triangles, see the inset on the left. The premedial lattice of the pyrochlore is the diamond lattice, which can be depleted by 1/4 of its bonds to obtain the hyperhoneycomb lattice. }
\label{fig:LatticeSummary2}
\end{figure}

Following our discussion in the main text relating the hyperoctagon and hyperkagome lattices as (pre)medial lattices of each other, one can
establish a similar set of relations for the hyperhoneycomb lattice as well. The medial lattice of the hyperhoneycomb is a lattice of corner-sharing triangles illustrated in Fig.~\ref{fig:3DKagome}, which can be considered another generalization of the kagome lattice to three spatial dimension, albeit one distinct from the hyperkagome and one which we dub {\em orthorhombic-kagome} lattice. Its main motif are sheets of two parallel triangle lines, which are staggered in a rotated way as illustrated in Fig.~\ref{fig:3DKagome}. 
The set of relations between the hyperhoneycomb and orthorhombic-kagome lattices as well as their relation to the pyrochlore and diamond lattices are summarized in Fig.~\ref{fig:LatticeSummary2}.

\subsection{Kitaev model}

Similar to the hyperoctagon lattice we can define a covering of $xx$, $yy$, and $zz$-couplings on the hyperhoneycomb lattice which is
commensurate with a four-site unit cell. This unit cell and related translation vectors as indicated in Fig.~\ref{fig:hyperhoneycomb_UnitCell}. 
The Kitaev Hamiltonian then takes the form 
\begin{align}
\label{eq:spin_Ham real space}
H=-\sum_{\mathbf{R}} & J_{x}\sigma_{1}^{x}(\mathbf{R})\sigma_{4}^{x}(\mathbf{R}-\mathbf{a}_{3})+J_{y}\sigma_{1}^{y}(\mathbf{R})\sigma_{4}^{y}(\mathbf{R}-\mathbf{a}_{3}+\mathbf{a}_{1})\nonumber \\
 & +J_{z}\sigma_{1}^{z}(\mathbf{R})\sigma_{2}^{z}(\mathbf{R})+J_{x}\sigma_{2}^{x}(\mathbf{R})\sigma_{3}^{x}(\mathbf{R})\nonumber \\
 & +J_{y}\sigma_{3}^{y}(\mathbf{R})\sigma_{2}^{y}(\mathbf{R}+\mathbf{a}_{2})+J_{z}\sigma_{3}^{z}(\mathbf{R})\sigma_{4}^{z}(\mathbf{R})\, ,
\end{align}
where $\mathbf R$  denotes the unit cell position. 

\begin{figure}[t]
\includegraphics[width=\columnwidth]{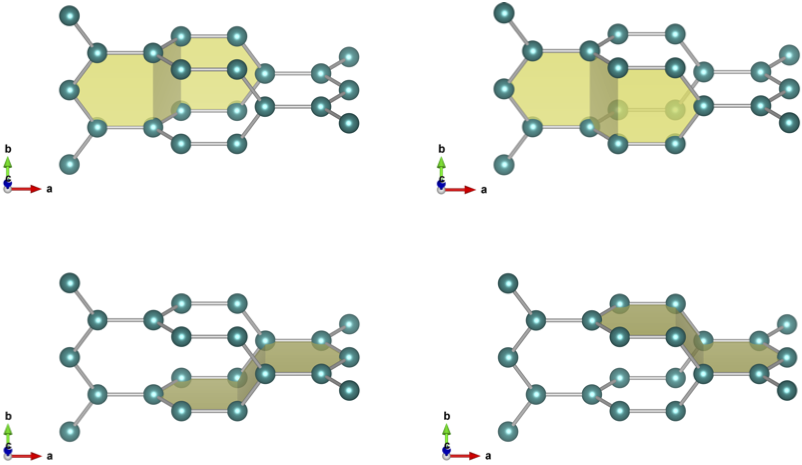}
\caption{(color online) The minimal closed surface is spanned by four loops in the hyperhoneycomb lattice.}
\label{fig:Loops_hyperhoneycomb}
\end{figure}

Similar to our analysis of the Kitaev model on the hyperoctagon lattice we can identify conservative quantities for this model by considering the structure of closed loops, which again have length ten for this model. For each elementary loop one can again identify a conserved quantity via the loop operators $W_l$ \eqref{eq:loopOperator}, which again have eigenvalues $\pm 1$. In contrast to the hyperoctagon lattice, the smallest volume enclosed by these elementary loops is now formed by {\em four} loops as illustrated in Fig.~\ref{fig:Loops_hyperhoneycomb}. Graphically speaking this constrains each `tetraeder' to have an even number of loops with eigenvalue $-1$. A counting argument similar to the one presented in our main analysis in Section \ref{sec:flux_sectors} shows that there are $2^{2N}$ different flux sectors, where $N$ is the number of unit cells. Numerical simulations \cite{Mandal09, hyperKimchi} indicate that the ground state indeed resides in the zero flux sector, which is why we restrict the following discussion to this sector.  

\subsection{Majorana spectrum and Fermi surface}

In order to solve the Hamiltonian, we proceed as in the main text and introduce four types of Majorana fermions per site to write $\sigma_j^\gamma(\mathbf R)=i a_j^\gamma(\mathbf R)c_j(\mathbf R)$.
Introducing bond operators $\hat u_{ij}=i a_i ^\gamma a_j ^\gamma$ and choosing the zero-flux sector as a reference sector, we again obtain a free fermion Hamiltonian of Majorana fermions hopping in a static $\mathbb Z_2$ gauge field
\begin{align}
H=i\sum_{\mathbf{R}} & J_{x}c_{1}(\mathbf{R})c_{4}(\mathbf{R}-\mathbf{a}_{3})+J_{y}c_{1}(\mathbf{R})c_{4}(\mathbf{R}-\mathbf{a}_{3}+\mathbf{a}_{1})\nonumber \\
 & +J_{z}c_{1}(\mathbf{R})c_{2}(\mathbf{R})+J_{x}c_{2}(\mathbf{R})c_{3}(\mathbf{R})\nonumber \\
 & +J_{y}c_{3}(\mathbf{R})c_{2}(\mathbf{R}+\mathbf{a}_{2})+J_{z}c_{3}(\mathbf{R})c_{4}(\mathbf{R})\:.\label{eq:majorana_Ham real space}
\end{align}
After a Fourier transformation \eqref{eq:FT}, the Hamiltonian is straightforward to diagonalize. 
The principle energy bands in the Majorana spectrum are thereby found to be
\begin{multline}
\label{eq:hh_energy}
E(\mathbf{k})  = \pm\frac{1}{2\sqrt{2}}\times \\ 
\sqrt{\Delta_{\mathbf{k}}\pm\sqrt{\Delta_{\mathbf{k}}^{2}-4[(J_{z}^{2}-|\delta{}_{1}||\delta{}_{2}|)^{2}+2J_{z}(1-\cos\phi_{k})|\delta_{1}||\delta_{2}|]}}\,,\\
\end{multline}
 where 
\begin{align}
\label{eq:def_of_quantities}
\Delta_{\mathbf{k}} & =(|\delta_{1}|^{2}+|\delta_{2}|^{2}+2J_{z}^{2}) \,,\nonumber \\
\delta_{1} & =J_{x}+e^{2\pi ik_{1}}J_{y} \,, \nonumber \\
\delta_{2} & =J_{x}+e^{2\pi ik_{2}}J_{y} \,, \nonumber \\
e^{i\phi_{k}} & =e^{-2\pi ik_{3}}\frac{\delta_{1}\delta_{2}}{|\delta_{1}||\delta_{2}|}\, .
\end{align}
The momenta $k_1,\ldots ,k_3$ are defined as the coefficients of the reciprocal lattice vectors, i.e. $\mathbf k=\sum_j k_j \mathbf q_j $ with $\mathbf q_i \cdot \mathbf a_j =2\pi\delta_{i,j}$. 

\begin{figure}[t]
\includegraphics[width=\columnwidth]{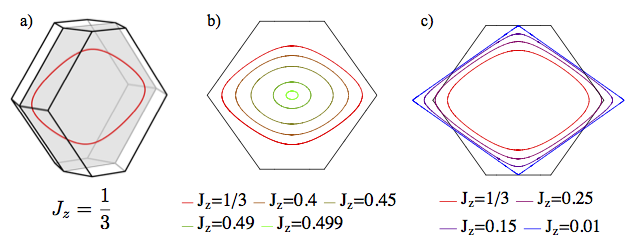}
\caption{(color online) 
Gapless less in the Kitaev model on the hyperhoneycomb lattice. Panel a) shows the position of the gapless line in the Brillouine zone at the isotropic point $J_x=J_y=J_z$. The gapless line is located in the plane $k_x+k_y=0$, which is indicated in grey. The other panels show the behavior of the gapless line when a) increasing $J_x$ and b) decreasing $J_x$, while setting $J_y=J_z=(1-J_x)/2$. The extent of the Brillouine zone in the plane $k_x+k_y=0$ is indicated by the hexagon.  }
\label{fig:hyperhoneycomb_gapless}
\end{figure}

Zero-energy solutions are obtained by setting the second term in the root \eqref{eq:hh_energy} to zero, which implies
\begin{align}
\label{eq:hh_constraint}
J_{z}^{2} & =|\delta_{1}||\delta_{2}|\nonumber \\
e^{i\phi_{k}} & =1. 
\end{align}
 The first line of Eq. (\ref{eq:hh_constraint}) can be inverted to yield
\begin{align}
\cos 2\pi  k_{2} & =\frac{J_{z}^{4}-(J_{x}^{2}+J_{y}^{2})(J_{x}^{2}+J_{y}^{2}+2J_{x}J_{y}\cos 2\pi  k_{1})}{2J_{x}J_{y}(J_{x}^{2}+J_{y}^{2}+2J_{x}J_{y}\cos 2\pi  k_{1})}\:,
\end{align}
 the second line determines the (unique) value of $k_{3}$ given $k_{1}$
and $k_{2}$.  
The gapless Majorana modes thus form a {\em line} in momentum space with linear dispersion along the normal directions. The line of gapless modes and its dependence on the coupling constant $J_z$ -- setting $J_x=J_y=(1-J_z)/2$ --  is shown in Fig~\ref{fig:hyperhoneycomb_gapless}.  For this choice of parameters the gapless line always lies in the $k_x+k_y=0$ plane, although this is no longer true when $J_x \neq J_y$.  We note that when approaching the gapped phase at $J_z=1/2$, $J_x=J_y=1/4$, the gapless line shrinks to a point at $(0,0,0)$.


\begin{thebibliography}{99}
\bibitem{HyperoctagonFiles}
For the convenience of the inclined reader we provide not only a VESTA visualization file of the hyperoctagon lattice, but also a 3D printing file in the auxiliary material of this
arXiv submission.
\bibitem{FootnoteOtherLattices}
Further generalizations of the Kitaev models also to lattices with higher vertex coordination number, in particular the kagome, triangular, hyperkagome and pyrochlore lattices have been studied \cite{KleinDuality2,Daghofer12,Becker14}.
It should however be noted that it is precisely the higher coordination number of the vertices in these lattices which prohibits to follow the same analytical route that can be used for the trivalent ones.
\bibitem{FootnoteZeroes} The zeroes on the diagonal are, in fact, not important for the argument.
\end{thebibliography}
\end{document}